\documentclass[twocolumn]{aastex631}
\usepackage{hyperref,amsmath}

\newcommand{\REV}[1]{{ {#1}}}

\newcommand\g{\; {\rm g}}
\newcommand\s{\; {\rm s}}
\newcommand\pcc{\;{\rm cm}^{-3}}
\newcommand\Msun{\; {\rm M}_{\odot}}
\newcommand\kms{\; {\rm km}\,{\rm s}^{-1}}

\newcommand\Myr{\;{\rm Myr}}

\newcommand\pc{\;{\rm pc}}

\newcommand\sfrunit{\Msun \pc^{-2} \Myr^{-1}}

\newcommand\Punit{\pcc\,{\rm K}}
\newcommand\surfunit{\Msun\,{\rm pc^{-2}}}
\newcommand\rhounit{\Msun\,{\rm pc^{-3}}}
\newcommand\K{\;{\rm K}}

\newcommand\pderiv[2]{\frac{\partial {#1}}{\partial {#2}}}

\newcommand{\tdep}{t_{\rm dep}}
\newcommand{\tdyn}{t_{\rm dyn}}
\newcommand{\tff}{t_{\rm ff}}

\newcommand{\seff}{\sigma_{\rm eff}}

\newcommand{\Ptot}{P_{\rm tot}}
\newcommand{\Pth}{P_{\rm th}}
\newcommand{\Pturb}{P_{\rm turb}}
\newcommand{\Pimag}{\Pi_{\rm mag}}

\newcommand{\Hg}{H_{\rm g}}

\newcommand{\Upstot}{\Upsilon_{\rm tot}}
\newcommand{\Upsth}{\Upsilon_{\rm th}}

\newcommand{\SSFR}{\Sigma_\mathrm{SFR}}
\newcommand{\Sg}{\Sigma_\mathrm{g}}
\newcommand{\rhog}{\rho_\mathrm{g}}

\defcitealias{2022ApJ...936..137O}{OK22}
\defcitealias{Springel:2003}{SH03}

\shorttitle{Towards PRFM Implementation in Cosmological Simulations}
\shortauthors{Hassan et al}

\begin{document}

\title{
Towards Implementation of the Pressure-Regulated, Feedback-Modulated Model of Star Formation in Cosmological Simulations: Methods and Application to TNG
}

\author[0000-0002-1050-7572]{Sultan Hassan}
\altaffiliation{NHFP Hubble fellow}
\affiliation{Center for Cosmology and Particle Physics, Department of Physics, New York University, 726 Broadway, New York, NY 10003, USA}
\affiliation{Center for Computational Astrophysics, Flatiron Institute, 162 5th Ave, New York, NY 10010, USA}\affiliation{Department of Physics \& Astronomy, University of the Western Cape, Cape Town 7535,
South Africa}

\author[0000-0002-0509-9113]{Eve C. Ostriker}
\affiliation{Department of Astrophysical Sciences, Princeton University, Princeton, NJ 08544, USA}
\affiliation{Institute for Advanced Study, 1 Einstein Drive, Princeton, NJ 08540, USA}

\author[0000-0003-2896-3725]{Chang-Goo Kim}
\affiliation{Department of Astrophysical Sciences, Princeton University, Princeton, NJ 08544, USA}

\author[0000-0003-2630-9228]{Greg L. Bryan}
\affiliation{Department of Astronomy, Columbia University, 550 W 120th Street, New York, NY 10027, USA}
\affiliation{Center for Computational Astrophysics, Flatiron Institute, 162 5th Ave, New York, NY 10010, USA}

\author[0000-0001-8293-3709]{Jan D. Burger}
\affiliation{Max-Planck-Institut f{\"u}r Astrophysik, Karl-Schwarzschild-Str. 1, D-85748, Garching, Germany}

\author[0000-0003-3806-8548]{Drummond B. Fielding}
\affiliation{Center for Computational Astrophysics, Flatiron Institute, 162 5th Ave, New York, NY 10010, USA}
\affiliation{Department of Astronomy, Cornell University, Ithaca, NY 14853, USA}

\author[0000-0002-1975-4449]{John C. Forbes}
\affiliation{Center for Computational Astrophysics, Flatiron Institute, 162 5th Ave, New York, NY 10010, USA}
\affiliation{School of Physical and Chemical Sciences|Te Kura Mat\={u}, University of Canterbury,
Private Bag 4800, Christchurch 8140,
New Zealand}

\author[0000-0002-3185-1540]{Shy Genel}
\affiliation{Center for Computational Astrophysics, Flatiron Institute, 162 5th Ave, New York, NY 10010, USA}
\affiliation{Columbia Astrophysics Laboratory, Columbia University, 550 West 120th Street, New York, NY 10027, USA}

\author[0000-0001-6950-1629]{Lars Hernquist}
\affiliation{Center for Astrophysics, Harvard \& Smithsonian, 60 Garden Street, Cambridge MA, USA}

\author[0000-0002-4232-0200]{Sarah M. R. Jeffreson}
\affiliation{Center for Astrophysics, Harvard \& Smithsonian, 60 Garden Street, Cambridge MA, USA}

\author[0000-0002-0045-5684]{Bhawna Motwani}
\affiliation{Department of Astronomy, Columbia University, 550 W 120th Street, New York, NY 10027, USA}

\author[0000-0002-9849-877X]{Matthew C. Smith}
\affiliation{Max-Planck-Institut f{\"u}r Astrophysik, Karl-Schwarzschild-Str. 1, D-85748, Garching, Germany}

\author[0000-0002-6748-6821]{Rachel S. Somerville}
\affiliation{Center for Computational Astrophysics, Flatiron Institute, 162 5th Ave, New York, NY 10010, USA}

\author[0000-0001-8867-5026]{Ulrich P. Steinwandel}
\affiliation{Center for Computational Astrophysics, Flatiron Institute, 162 5th Ave, New York, NY 10010, USA}

\author[0000-0001-7689-0933]{Romain Teyssier}
\affiliation{Department of Astrophysical Sciences, Princeton University, Princeton, NJ 08544, USA}
\affiliation{Program in Applied and Computational Mathematics, Princeton University, Fine Hall Washington Road, Princeton NJ08544-1000 USA}

\correspondingauthor{Sultan Hassan}
\email{sultan.hassan@nyu.edu, eco@astro.princeton.edu, cgkim@astro.princeton.edu}

\begin{abstract}
Traditional star formation subgrid models implemented in cosmological galaxy formation simulations, such as that of \citet[][hereafter \citetalias{Springel:2003}]{Springel:2003}, employ adjustable parameters to satisfy constraints measured in the local Universe. In recent years, however, theory and spatially-resolved simulations of the turbulent, multiphase, star-forming ISM have begun to produce new
first-principles models, which when fully developed can replace traditional subgrid prescriptions. This approach has advantages of being physically motivated and predictive rather than empirically tuned, and allowing for varying environmental conditions rather than being tied to local Universe conditions. As a prototype of this new approach, by combining calibrations from the TIGRESS numerical framework with the Pressure-Regulated Feedback-Modulated (PRFM) theory, simple formulae can be obtained for both the gas depletion time and an effective equation of state. Considering galaxies in TNG50, we compare the ``native" simulation outputs with post-processed predictions from PRFM. At TNG50 resolution, the total midplane pressure is nearly equal to the total ISM weight, indicating that galaxies in TNG50 are close to satisfying vertical equilibrium. The measured gas scale height is also close to theoretical equilibrium predictions. The slopes of the effective equations of states are similar, but with effective velocity dispersion normalization from SH03 slightly larger than that from current TIGRESS simulations. Because of this and the decrease in PRFM feedback yield at high pressure, the PRFM model predicts shorter gas depletion times than the \citetalias{Springel:2003} model at high densities and redshift. Our results represent a first step towards implementing new, numerically calibrated subgrid algorithms in cosmological galaxy formation simulations.
\end{abstract}


\section{Introduction} \label{sec:intro}

In the large-box cosmological simulations needed to study the history and statistics of galaxy formation, limitations of spatial resolution make it impossible to directly follow the dynamics and thermodynamics of the multiphase interstellar medium (ISM), and  the physical processes involved in star formation and stellar feedback.  As a result, simple prescriptions for star formation must be adopted, which are generally applied when the density exceeds a threshold designated to select star-forming regions of galaxies.  

For $m_\mathrm{g}$ the gas mass in a cell or particle that is above the density threshold in the cosmological simulation, the most common prescription for star formation is to set $\dot{m}_*/m_\mathrm{g} = t_{\rm dep,0}^{-1} (n_{\rm H}/n_{\rm th})^{1/2}$, where $n_{\rm H}=\rhog/(\mu_{\rm H} m_p)$ is the local measured gas number density, $n_{\rm th}$ is an adopted threshold density, and $t_{\rm dep,0}$ is a gas depletion time at density $n_{\rm th}$.  The functional form here is motivated by the physical concept of making the specific star formation rate (SFR)  inversely proportional to the free-fall time in the gas,  $t_{\rm ff}=(3\pi/32 G\rhog)^{-1/2}$.  
Following the star formation and ISM model introduced by \citet{Springel:2003} (hereafter \citetalias{Springel:2003}) with some minor modifications, this approach is employed in Illustris  \citep{2013MNRAS.436.3031V}, where $t_{\rm dep,0}=2.2$ Gyr and 
$n_{\rm th}=0.13 \pcc$ are adopted (see \autoref{sec:TNG50}).
The overall normalization of the SFR is sometimes also cast in terms of an efficiency per free-fall time, $\varepsilon_{\rm ff} = t_{\rm ff}(n_{\rm th})/t_{\rm dep,0}$; the parameter choices in Illustris correspond to $\varepsilon_{\rm ff} =0.06$.  
Other cosmological simulation frameworks adopt similar practices.  In the MUFASA and SIMBA simulations, star formation follows the above scaling with density, but only in gas that is designated as molecular, based on a subgrid model \citep{2016MNRAS.462.3265D,2019MNRAS.486.2827D}.  In the EAGLE simulations \citep{2015MNRAS.446..521S}, $\dot{m}_*/m_\mathrm{g}$ is explicitly set to be proportional to a power law in pressure, rather than density, although with the effective equation of state (eEoS) $P\propto \rhog^{4/3}$ that is adopted above $n_{\rm H}=0.1\pcc$, the resulting relation is $\dot{m}_*/m_\mathrm{g} \propto n_{\rm H}^{0.3}$; the coefficient in the specific SFR is empirically set based on \citet{Kennicutt:1998}.\footnote{The EAGLE approach to implementing star formation via a power law in pressure is motivated by \citet{2008MNRAS.383.1210S} as a way to directly reproduce empirical Kennicutt-Schmidt relations in which $\SSFR$ follows a power law in $\Sg$.  Two practical difficulties with this, however, are that stellar gravity is in general as important as gas gravity in setting the equilibrium pressure, and that the pressure in a simulation will be lower than it should be if the spatial resolution is insufficient (see \autoref{sec:equil}).} In general, normalizations adopted in empirically-constrained star formation prescriptions are based on low-redshift observations.

More recently, other subgrid models of star formation have been implemented in cosmological zoom and isolated-galaxy simulations that are motivated by turbulence-driven theories developed for conditions within GMCs \citep[e.g.][]{Semenov2016,Kimm2017,Kretschmer2020,2020MNRAS.495..199G,Nunez2021,Dubois2021,2023arXiv230110251J}, or employ subgrid models for molecular content \citep[e.g.][]{Gnedin2009,Hopkins2018,Feldmann2023}. 
A stochastic star formation rate scaling with local density $\dot{m}_*/m_{\rm g}\propto \tff^{-1} \propto \rhog^{1/2}$ is sometimes also adopted in dwarf galaxy simulations with even higher resolution that employ additional physics, including direct feedback \citep[e.g.][]{2016MNRAS.458.3528H,2019MNRAS.482.1304E,2020ApJ...891....2L,2021MNRAS.506.3882S,2022arXiv221203898S, 2023MNRAS.526.1408S}. Interestingly, simulations employing Lagrangian and Eulerian methods can have very different results because in the former case collapse continues to much smaller scales, resulting in much higher net star formation efficiency, and greater feedback \citep{Hu2023}.   
Cosmological zoom simulations that resolve high densities have much higher thresholds than large-box simulations; e.g FIRE-2 adopts a default density threshold of $n_\mathrm{th}=10^3\pcc$ \citep{Hopkins2018}.

The parameters chosen for cosmological subgrid models are set based on empirically-determined gas depletion times $\tdep \equiv M_\mathrm{g}/\dot M_*$ (note that we use upper-case ``M'' here to distinguish empirical measurements from the lower-case ``m'' representing a mass element in a numerical simulation).
Observationally, $\tdep \sim 1$ Gyr for the majority of gas in the local Universe, although depletion times  drop in high density regions of the ISM (notably galactic centers)  and at higher redshift \citep[e.g.][]{1998ApJ...498..541K,2010MNRAS.407.2091G,2012ARA&A..50..531K,2017ApJ...849...26U,2019ApJ...882....5W,2020ARA&A..58..157T}.  With the majority of star-forming gas in cosmological simulations just over the density threshold, the adopted $\tdep$ normalizations give reasonable agreement with the overall observed conversion rate from gas to stars.

It is not clear, however, that the scaling relation $\dot{m}_*/m_\mathrm{g} \propto t_{\rm ff}^{-1} \propto \rhog^{1/2}$ is justified in the regimes of density accessible to cosmological simulations. Within giant molecular clouds (GMCs, at densities $\gtrsim 10^2 \pcc$ and temperatures $\sim 10 \K$),  fragmentation to form stars is regulated by the interactions among gravity, supersonic turbulence, and magnetic fields. The specific SFR under GMC conditions is often characterized as $\dot M_*/M_{\rm g} = \varepsilon_{\rm ff} t_{\rm ff}^{-1}$, but from theory and simulations, $\varepsilon_{\rm ff}$ is in fact quite sensitive to the virial parameter (the ratio of twice kinetic to gravitational energy) and mass-to-magnetic flux ratio \citep[e.g.][see also reviews by \citealt{FederrathKlessen2012,Padoan+2014} for further discussion of turbulence-controlled theories of star formation on GMC scales]
{krumholz05,padoan12,2021ApJ...911..128K}. The virial parameter and mass-to-flux ratio depend  on conditions on larger scales and on the GMC formation process, and based on observations appear to vary considerably \citep[e.g.][]{2021ApJ...920..126E}.  Empirically, the mean value is $\varepsilon_{\rm ff}\approx 0.01$ for molecular gas in GMCs in local-Universe galaxies, but there is an order of magnitude variation about this \citep[e.g.][]{2012ApJ...745...69K,2018ApJ...861L..18U,2022ApJ...929L..18E,2023ApJ...945L..19S}.
Meanwhile, the efficiency of star formation over a cloud lifetime depends on the action of feedback \citep[see e.g.][and references therein]{2023ASPC..534....1C}.  
Moreover, GMCs are overdense compared to average ISM conditions, and they (and the processes that form them) cannot be resolved in cosmological simulations. 

The densities that {are} accessible in large-box cosmological galaxy formation simulations 
represent averages over the whole of the multiphase ISM, rather than conditions similar to those in GMCs. When averaged over very large scales and multiple ISM phases in galaxies, it is known that SFRs respond to a highly complex array of physical processes rather than just the mean free-fall time in the gas \citep[see e.g.~reviews of][]{2007ARA&A..45..565M,2020SSRv..216...68G,2023ASPC..534....1C,2020SSRv..216...64K}.  From a theoretical point of view, it has proven fruitful conceptually to focus on the requirements for thermal and dynamical equilibrium in the mass-containing components of the ISM, and the role of star formation feedback in maintaining this \citep{2010ApJ...721..975O,2011ApJ...731...41O,2011ApJ...743...25K}.  The pressure-regulated, feedback-modulated (PRFM) theory of the star-forming ISM \citep[summarized in][hereafter \citetalias{2022ApJ...936..137O}]{2022ApJ...936..137O} formalizes these ideas and emphasizes the importance of the feedback ``yield,'' defined as the ratio between total gas pressure and star formation rate per unit area, $\SSFR$.

Recent advances in theoretical and computational modeling of the star-forming ISM provide an opportunity to develop more sophisticated subgrid model treatments in cosmological simulations.  A key goal of the Learning the Universe Simons Collaboration\footnote{https://www.learning-the-universe.org/} -- as well as one of its precursors, the SMAUG collaboration
-- is to replace the current, empirically-calibrated prescriptions for star formation with new subgrid models that are instead calibrated from ``full-physics'' ISM simulations, such as that implemented in the ``TIGRESS'' and ``TIGRESS-NCR'' frameworks \citep{2017ApJ...846..133K,2023ApJS..264...10K,2023ApJ...946....3K,2024arXiv240519227K} and their successors.\footnote{Current simulations with the TIGRESS-NCR framework include supernova feedback, UV radiative transfer via adaptive ray-tracing, and photochemistry/nonequilbrium cooling for hydrogen and key carbon and oxygen species, while future work will include a cosmic ray fluid as implemented based on \citet{2021ApJ...922...11A}.} In order to develop a new star formation subgrid model that can be used for galaxy formation/evolution, it is necessary to conduct parameter survey simulations at $\sim \pc$ resolution over a range of conditions in galaxies, and to fit the resulting SFRs \citep[as begun in][]{2013ApJ...776....1K,2020ApJ...900...61K,2022ApJ...936..137O}.  The functional forms adopted should be motivated both by fundamental theoretical considerations of ISM physics 
and by knowledge of what parameters are available in cosmological simulations, \REV{and robust to changes in resolution}.  In this paper, we present a first demonstration of applying a new subgrid star formation model -- based on the PRFM theory and calibrated against high-resolution ISM simulations -- to galaxies as formed in much lower resolution cosmological simulations.  

As necessitated by limited resolution in cosmological simulations, in addition to a subgrid star formation model it is typical practice to adopt an eEoS for gas above some density threshold.  This serves an important numerical purpose of suppressing self-gravitating fragmentation that could otherwise occur \citep[e.g.][]{1997ApJ...489L.179T}.  Additionally, an eEoS can in principle represent a physical relationship between the mean effective pressure and the mean density in the ISM.  Given the complex, multiphase nature of the ISM gas, deriving relationships of this kind is nontrivial.  In \citetalias{Springel:2003}, an eEoS was proposed based on a theoretical model of the ISM (see \autoref{sec:TNG_SFR} for more details). As noted above, EAGLE adopts $P\propto \rho^{4/3}$ for star-forming gas \citep{2015MNRAS.446..521S}, and the same is true for MUFASA and SIMBA \citep{2016MNRAS.462.3265D,2019MNRAS.486.2827D}.   
With the recent development of high resolution ISM simulations that resolve multiphase gas, including radiative transfer and chemistry as needed for following the main heating and cooling processes, and resolving collapse leading to star formation as well as radiation and supernova feedback, it is now possible to instead \emph{calibrate} an eEoS. eEoS functions calibrated in this way can then serve as a subgrid ISM model in cosmological simulations. Here, we provide a first demonstration of applying a calibrated eEoS function -- based on a set of TIGRESS simulations --  to compute the pressure-density relationship in cosmological simulations of galaxies.

In addition to the effects of star formation feedback in pressurizing ISM gas, which can be captured for the purpose of a cosmological subgrid model via an eEoS, feedback also leads to driving of galactic winds.  As part of the SMAUG and Learning the Universe Simons Collaboration, we are additionally developing new subgrid approaches to modeling wind driving that take into account the essential multiphase nature of winds.
\REV{This ``Arkenstone'' framework has  (interacting) hot and cool phases that require separate implementations \citep{2024MNRAS.527.1216S};  outflow loading factors for each component are calibrated from the same  high-resolution TIGRESS simulations  as are used for star formation and eEoS models  \citep{2020ApJ...900...61K,2020ApJ...903L..34K}.
}

This paper is organized as follows: We first discuss TNG50 in \S\ref{sec:TNG50}, including presenting a detailed review of methods from \citetalias{Springel:2003} in 
\S\ref{sec:TNG_SFR}, and describing measurement of local ISM properties from simulations in \S\ref{sec:localprob}. Formulae for quantifying ISM properties (including pressure, disk scale height, and the dynamical timescale) under vertical equilibrium are presented in \S\ref{sec:equil}, and the PRFM theory and TIGRESS resolved ISM simulations are reviewed in \S\ref{sec:prfm}. We then present a detailed comparison between quantities in TNG50 galaxies under the ``native'' \citetalias{Springel:2003} prescriptions, and under alternative prescriptions based on PRFM theory and TIGRESS simulations, considering the effective velocity dispersion, gas scale height, eEoS, depletion time, and 
observed star formation scaling
relations in \S\ref{sec:results}. Finally, we  summarize our key findings and discuss future applications in \S\ref{sec:conc}.

\section{TNG50 Simulations}\label{sec:TNG50}
The TNG50 simulation~\citep{2019MNRAS.490.3234N,2019MNRAS.490.3196P} is the highest resolution run of the IllustrisTNG simulation project, achieving a mass resolution approaching that of zoom-in simulations.  TNG50 evolves 
dark matter, gas, stars, and black holes within a simulation box of size 51.7$^3$ comoving Mpc$^3$, using 2160$^{3}$ each gas fluid elements and dark matter particles. The mean baryon and dark matter mass resolutions are 8.5$\times 10^{4}\, M_{\odot}$ and 4.5$\times 10^{5}\, M_{\odot}$, respectively. The minimum adaptive gravitational softening length for gas cells (comoving Plummer equivalent), and the $z=0$ softening of the collisionless components are 74$\pc$ and 288$\pc$, respectively (softenings for the collisionless component are smaller at $z>1$ since they are comoving).  Details of IllustrisTNG's implementation of various aspects of sub-grid physics involved in galaxy formation are covered in \citet{2017MNRAS.465.3291W, 
2018MNRAS.475..676S, 2018MNRAS.475..648P, 2018MNRAS.480.5113M, 2018MNRAS.477.1206N, 2018MNRAS.475..624N, 2018MNRAS.473.4077P}.
Given the importance of the star formation sub-grid model in the present work, we next describe it in somewhat more detail here.

\subsection{Star Formation and Effective Equation of State in TNG50}\label{sec:TNG_SFR}

In \citetalias{Springel:2003}, 
the SFR model is:
\begin{equation}\label{eq:SFR_SH03}
\dot{m}_*= (1-\beta)\frac{m_{\rm g}}
{t_{*}} \equiv \frac{m_{\rm g}}{\tdep}\, ,
\end{equation}
where $m_{\rm g}$
is the mass of gas 
eligible for star formation (gas particles with a hydrogen number density of $n_{\rm H} \geq 0.13 \rm \, cm^{-3}$), $t_{*}$ is a characteristic timescale (depending on the gas density) to convert this gas into stars with mass $m_*$, 
$\beta$ ($\approx 0.1$) is the mass fraction of stars that explode as supernovae and would be returned to the ISM very quickly, and $\tdep$ is the net instantaneous gas depletion time. The Illustris/TNG simulation approach \citep{2013MNRAS.436.3031V} modifies this slightly by explicitly including stellar mass return separately and therefore omitting the $(1-\beta)$ factor.

As discussed in \autoref{sec:intro}, Illustris and IllustrisTNG adopt the \citetalias{Springel:2003} prescription in which the star formation timescale $t_*$ follows a free-fall scaling with gas density.
The TNG50 simulation modifies this to allow for more efficient star formation at high densities (steeper dependence of SFR on $\rho$), in order to avoid a very short numerical time step \citep{2019MNRAS.490.3234N}.
In the modified SFR prescription,
\begin{equation}
t_*(\rho) \, = \,
\left\{\!\begin{aligned}
  \, &t_{*,0}\left(\frac{\rho}{\rho_{\rm th}}\right)^{-1/2}\, \quad &; \quad \rho\leq230\, \rho_{\rm th}\\[1ex]
  \, &t_*(230 \rho_{\rm th})\left(\frac{\rho}{230\rho_{\rm th}}\right)^{-1}\, \quad &; \quad \rho>230\, \rho_{\rm th}
\end{aligned}\right\} ,
\end{equation}
where $\rho=230 \rho_{\rm th}$ is equivalent to $n_{\rm H}=24.4\pcc$.  
Here $t_{*,0}\equiv t_{\rm dep,0}$ in the terminology of \autoref{sec:intro}.
In order to approximately reproduce the global empirical star formation relation of \citet{Kennicutt:1998}, \citetalias{Springel:2003} adopt $t_{*,0}=2.1$~Gyr,  while $t_{*,0}=2.2$~Gyr is used for the normalization of Illustris and IllustrisTNG.

\begin{figure*}
    \centering
    \includegraphics[scale=0.52]{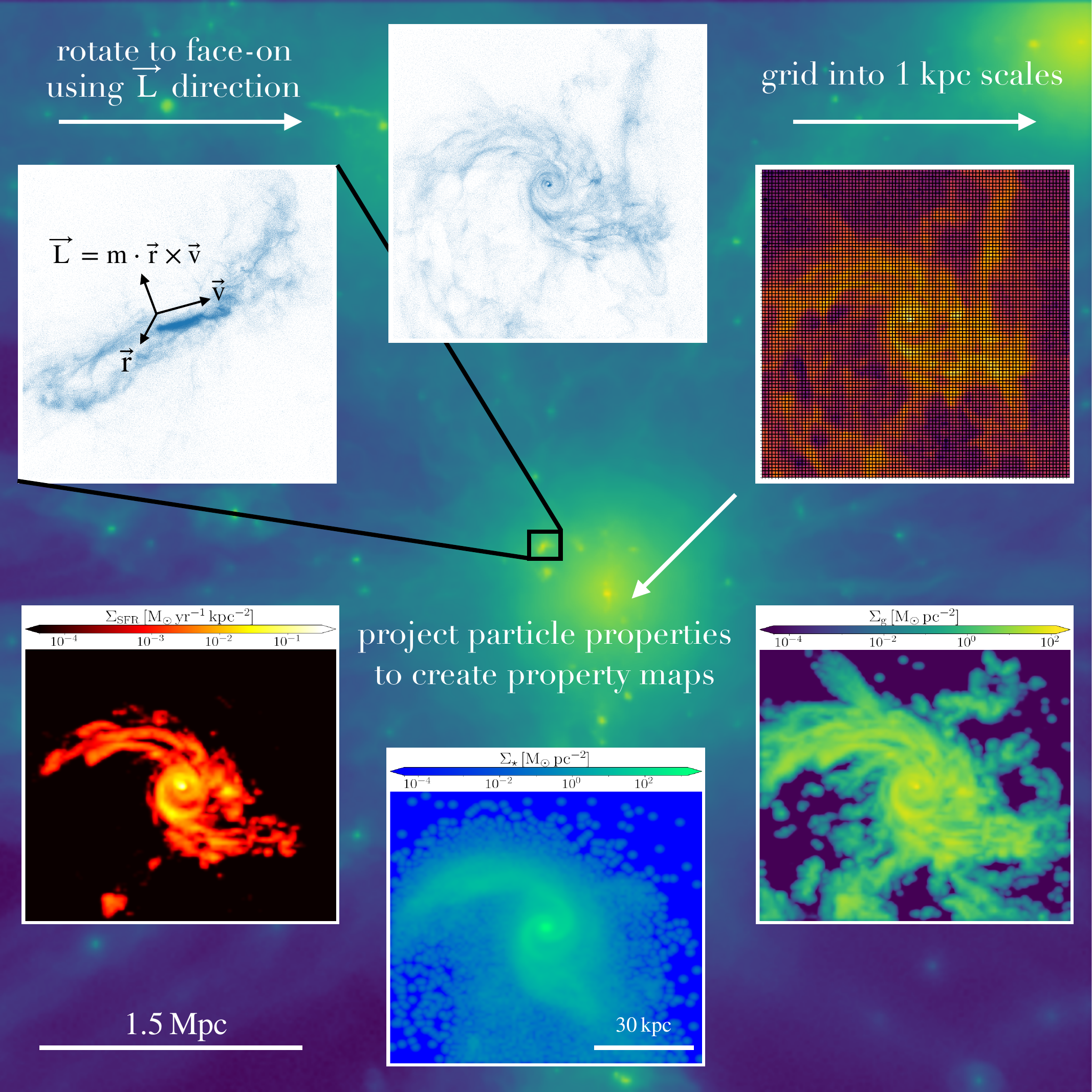}
    \caption{Schematic summary of the method used to measure the local galactic properties within 1 kpc scales for a random galaxy at z=0 with a stellar mass $M_{\star} = 3.2\times 10^{10}\, \rm M_{\odot}$. We start with the gas particle data to identify the angular momentum direction $\hat{L}$ (top left), and rotate the 3d coordinates and velocities to a face-on direction (top middle). We then create a grid with a pixel size of 1 kpc, and column heights of $\pm$ 0.1 and $\pm$ 10 kpc above/below the midplane (top right) to compute the midplane (e.g. pressure $P$) and integrated (e.g. surface density gas $\Sigma_{\rm g}$) properties, respectively. We finally create all property maps by projecting the relevant particle properties. We show below resulting maps of gas surface density ($\Sigma_{g}$, bottom right), stellar surface density ($\Sigma_{\star}$, bottom middle), and star formation rate density ($\Sigma_{\rm SFR}$, bottom left). We repeat the same procedure for all TNG50 galaxies at $z=0$ and $z=2$ within the mass range of $M_{\star}=10^{7-11}M_{\odot}$, and combine all maps to study the overall distributions of the local galactic properties in TNG50. These distributions will also be used to make predictions for the star formation using the PRFM model.}
    \label{fig:vis}
\end{figure*}

The IllustrisTNG simulation project makes use of the sub-grid multi-phase model developed by \citetalias{Springel:2003}. Inspired by \citet{McKee:1977} and \citet{Yepes1997}, the \citetalias{Springel:2003} model conceives of the ISM as consisting of a hot intercloud medium (comprising the majority of the ISM's energy density) within which is embedded a population of cold clouds (comprising the majority of the mass). 
It is assumed that the hot phase gains energy from supernovae \REV{and uses some of this energy in conductively evaporating cold clouds}; and then radiates away this energy \REV{(at the same average rate)}, leading to mass condensation into  cold clouds.  The coefficient in the condensation rate is adjusted so that the hot+cold subgrid model transitions to a single-temperature model at the point where the hot gas temperature drops to $10^5\K$, and the density threshold is chosen such that the temperature in the eEoS continuously connects at that density to an isothermal EoS at $10^4\K$.  
In the IllustrisTNG treatment of star-forming gas,  there is an interpolation between the temperature as predicted by the \citetalias{Springel:2003} eEoS, and an isothermal with $T=10^4\K$, with an interpolation parameter $q_{\rm EOS}=0.3$ \citep{2013MNRAS.436.3031V}. 

While the eEoS in IllustrisTNG, as adapted from \citetalias{Springel:2003}, is effective numerically and is physically motivated \REV{based on the original \citet{McKee:1977} three-phase theory}, ISM theory in recent years has taken a different perspective  
on the key physical processes controlling the state of the multiphase ISM, and the considerations needed to define an eEoS.  \REV{In particular, observations suggest that most of the hot ISM is at high enough temperature to be radiatively inefficient (with only a few percent of SN energy radiated in X-rays), producing orders of magnitude less emission in \ion{O}{6} than would be expected from the original \citet{McKee:1977} model that relied on cooling of the hot phase (\citealt{2005ARA&A..43..337C}; see also \citealt{2003ApJ...591..821O,2008ApJS..176...59B}). In modern ISM theory, SN energy partly goes into kinetic energy of cooler phases, partly is lost in a hot galactic wind, and partly is lost to post-shock cooling and cooling in turbulent radiative mixing layers at the edges of SN remnants, superbubbles, and in the galactic fountain; cooling directly from the hot phase is very limited \citep[e.g.][]{1993ApJ...407...83S,2010ApJ...719..523K,2015ApJ...802...99K,2017ApJ...834...25K,2019MNRAS.490.1961E,2020ApJ...894L..24F,2021MNRAS.508L..37T}.} 
From the point of view of characterizing the effective pressure in the ISM, thermal conduction is currently considered to play a subdominant role. While classical (Spitzer) thermal conduction can contribute to evaporation of cooler phases, increasing the mass of the hot phase and lowering its temperature \citep[][]{1977ApJ...211..135C,1977ApJ...218..377W}, the pressure of the hot medium in superbubbles is the same regardless of conduction, 
making the dynamical effect of supernovae on the cooler phases insensitive to conduction. 
Only if losses in turbulent radiative mixing layers were unphysically small would the evaporative mass flux be able to reduce the temperature of the hot medium significantly \citep[see e.g.~Eq. 45 of][]{2019MNRAS.490.1961E}, leading to enhanced cooling via \ion{O}{6} and other high-ionization lines. 
Thus, while the level of thermal conduction \citep[which tends to be suppressed by magnetic fields; e.g.~][]{2001ApJ...562L.129N,2008ApJ...678..274O,2021MNRAS.502.1263K} quantitatively affects the temperature and emission from the hot phase, the properties of the cooler phases that comprise most of the mass (and produce star formation) are not currently thought to be significantly affected by conduction.

Since the eEoS is intended to link a large-scale average gas density \REV{(equal to the surface density divided by twice the disk scale height)} with a large-scale average pressure, one must consider what physically determines the scale height of the warm ($T\sim 10^4\K$) and cold ($T\lesssim 100 \K$) gas that comprises the majority of the ISM mass. As we shall discuss in \autoref{sec:equil},
it is the turbulent, thermal, and magnetic pressures of the warm and cold gas itself that support these phases against gravity and determines their scale height. The pressure of the hot phase on large scales does not directly support cooler gas (this situation would be Rayleigh-Taylor unstable); rather, 
blast waves produced by supernovae shocks accelerate cooler phases, thereby setting their turbulent velocity dispersion and pressure. \REV{This turbulence acts in concert with shear to drive the galactic dynamo \citep{2015ApJ...815...67K}.  Meanwhile, the thermal pressure in warm and cold gas is controlled by far-UV radiation (plus cosmic ray) heating  proportional to the recent star formation rate.  Pressure unrelated to supernova energy injection was not directly considered in the \citetalias{Springel:2003} model.} 
Defining an eEoS \REV{consistent with modern ISM theory} requires calibrating the total effective velocity dispersion for cool gas phases (\REV{combining turbulent, magnetic, and thermal terms}), which since it depends on intricate details of feedback is best accomplished using resolved ISM simulations such as TIGRESS (see \autoref{sec:prfm}).

\begin{figure*}
    \centering
    \includegraphics[scale=0.65]{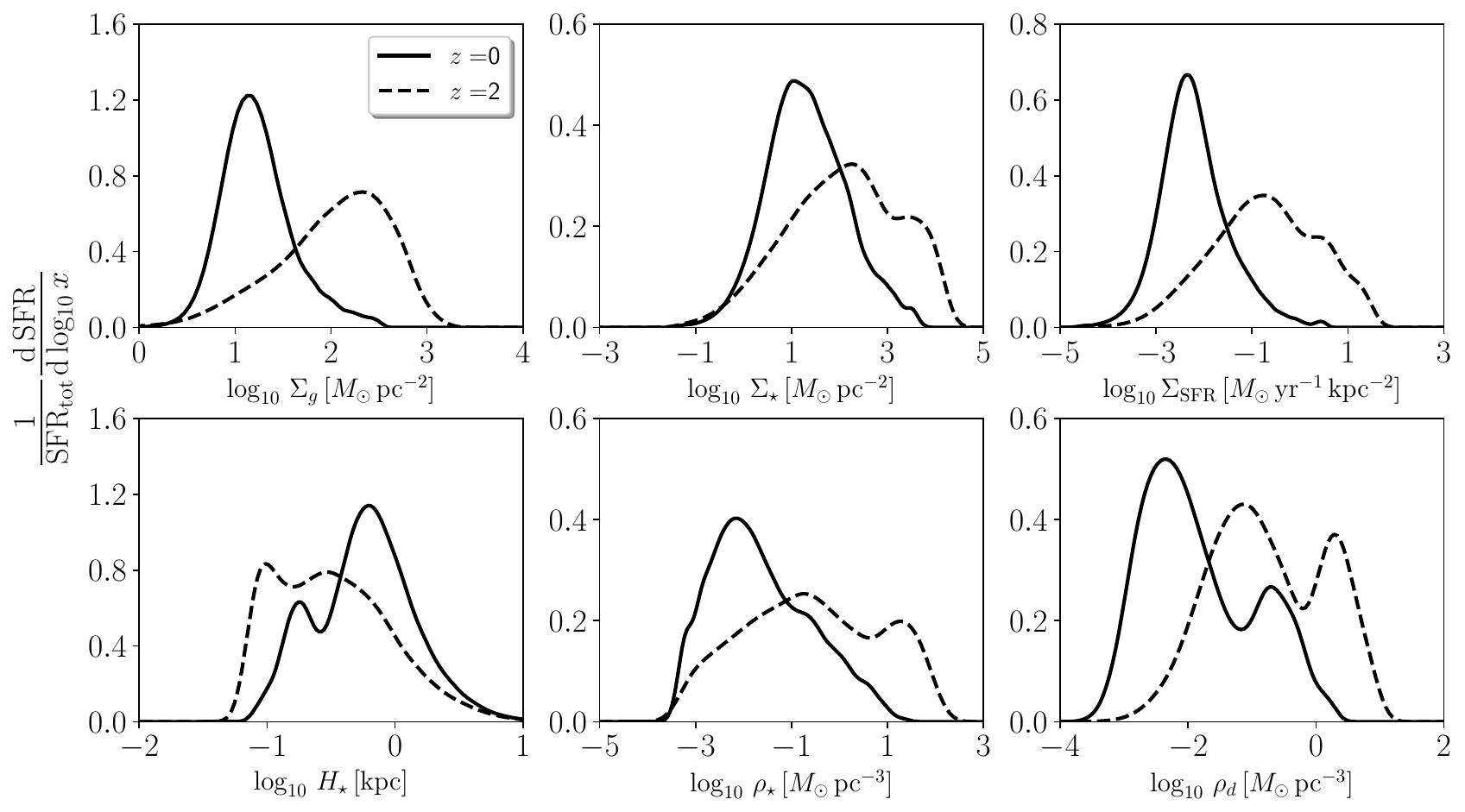}
    \caption{Distributions, weighted by contribution to the total $\rm SFR$ at each redshift,  of various local environmental properties from 
    all TNG50 galaxies at $z=0$ (solid) and $z=2$ (dashed). For each panel, we show  
    ${\rm SFR}_\mathrm{tot}^{-1} d\,{\rm SFR}/d\log_{10} {\it x}$,     
    where $x$ represents the property on the $x$-axis. Measurements are within 1 proper kpc pixels in maps projected parallel to the direction of the angular momentum. Top panels show surface density of gas ($\Sigma_{\rm g}$), stars ($\Sigma_{*}$), and SFR ($\Sigma_{\rm SFR}$); bottom panels show stellar scale height ($H_{*}$),  volumetric density of stars $\rho_{*}$, and dark matter $\rho_{\rm d}$. The distributions show a large decrease in characteristic densities from $z=2$ to $z=0$.}
    \label{fig:all_dist}
\end{figure*}

\subsection{Measurements of the Local Galactic Disk and Star Formation Properties in TNG50}\label{sec:localprob}
We follow closely the analysis presented in~\citet{Motwani:2022} and refer to it for extensive discussions. We here briefly describe how the different local properties are measured from the TNG50 simulation, and summarize the distributions of these properties.  We focus our presentation on properties at redshifts $z=0$ and $2$, but the trends shown hold more generally.\footnote{We have checked the results at $z=1,3$, and found a similar evolution can be predicted using $z=0,2$ results.} Due to  resolution limitations, we select central and satellite galaxies from TNG50 as follows: First, we consider only galaxies with stellar masses between 10$^{7-11}$ M$_{\odot}$. Second, we apply a minimum limit of 100 particles for all types (gas, stars, and dark matter), and a minimum SFR threshold of 5$\times$10$^{-4}$ M$_{\odot}/$yr. Third, we exclude any ``misidentified'' galaxies (according to the halo-finding criteria in TNG). Using these selection criteria, our final sample includes 10394 and 21039 galaxies at redshifts $z=0$ and $z=2$ respectively. 
\REV{The reason for adopting a minimum SFR threshold in our sample selection is that the PRFM theory may not apply to dynamically quenched galaxies with very low SFR, such as some ellipticals (e.g.~S. Jeffreson et al, 2024, in prep.).}

For each galaxy within our sample, we first use the angular momentum direction to rotate the spatial coordinates and velocities to a face-on view, with $\hat{z}$ along the angular momentum direction. Next, for each galaxy, we create a map with radius equal to twice the half stellar mass radius ($r^{\star}_{1/2}$) along the $x-y$ direction (map size = \{$-2 r^{\star}_{1/2}$, $+2 r^{\star}_{1/2}$\}).  Within each map, we assume a pixel size of 1  proper kpc, which is similar to the horizontal box size (and the averaging scale) in the TIGRESS star-forming ISM simulations. We use a column of $\pm$ 10  proper kpc along the $z$ direction to obtain projection maps of various properties, which include the gas surface density $\Sigma_{\rm g}$, stellar surface density $\Sigma_{*}$, the SFR surface density $\Sigma_{\rm SFR}$, stellar scale height $H_{*}$, 
stellar midplane volumetric density $\rho_{*}$, and dark matter volumetric density $\rho_{\rm d}$. For surface densities, we sum all masses within columns and divide them by the pixel area.  
The stellar scale height is defined as $H_*=\Sigma_*/(2\rho_*)$ with $\rho_*$ the midplane density measured by the mean stellar density within $z=\pm 100$ pc.
The dark matter volumetric density $\rho_{\rm d}$ is directly computed from the local total mass density around gas particles (snapshot-provided), which is estimated using the standard cubic-spline SPH kernel within a sphere enclosing the 64$\pm$1 nearest dark matter particles. 

For each map pixel, we also obtain several midplane properties of the gas for comparison with the PRFM prediction (\autoref{sec:results}), including the mass-weighted averages of pressure components (thermal, turbulent, and magnetic -- see \autoref{sec:results_equil} for definitions) 
and gas density $\rhog$. In particular, the turbulent pressure is an average of $\rho v_z^2$, where the vertical velocity $v_z$ is computed relative to the mean galactic velocity in the coordinate system where $\hat z$ is along the direction of the angular momentum. These measurements are used to calculate the effective velocity dispersion (\autoref{sec:results_seff}) and gas scale height (\autoref{sec:results_Hg}).
The ``midplane region'' is defined  as being within 100 proper pc above/below the $z=0$ plane\footnote{The adopted thickness of the midplane region is comparable to the softening length as quoted earlier. We have also checked using different thicknesses including $z=\pm0.2$, 0.5, and 1 kpc and the distributions remain largely unaffected.}. All galactic property maps are directly computed from the instantaneous snapshots at $z=0$ and $z=2$ in proper units. A schematic summary of how we measure these property maps is depicted in~\autoref{fig:vis} for a random spiral galaxy of stellar mass $M_{\star} = 3.2\times 10^{10} M_{\odot}$ at $z=0$.

It is worth noting that all comparisons performed in this work are at the level of 1 proper kpc pixels, from all galaxies combined. We intentionally do not present the analysis in terms of different galaxy masses or types, but rather focus on the broad distributions of the 1 proper kpc patches from different environments within the whole simulation domain at different cosmic times. This ensures that we uniformly sample all galaxy environments.

\autoref{fig:all_dist} shows the $\Sigma_{\rm SFR}$-weighted distributions of various local environmental properties from the TNG50 maps created as described above, for  
$z=0$ (solid) and $z=2$ (dashed). As expected, all surface ($\Sigma_{\rm g}, \Sigma_{*}, \Sigma_{\rm SFR}$) and volumetric  ($\rho_{*}, \rho_{\rm d}$) densities are higher at high redshifts (roughly 1 -- 2 orders of magnitude higher at $z=2$ compared to $z=0$). The stellar scale height $H_{*}$ distributions show the expected opposite evolution,  with $H_{*}$ 
lower at $z=2$ compared to $z=0$; the redshift evolution of $H_*$ is weaker than other quantities, however.

Both the stellar and dark matter distributions show a double-peak profile owing to different galactic regions (inner/outer) and types. See \citet{Motwani:2022} for a more detailed discussion of the multi-variate distributions of these properties and others.
In particular, the peak at higher  $\rho_d$ is associated with low-mass galaxies, whereas the more prominent peak at lower $\rho_d$  is close to the peak at $\rho_d \sim 10^{-2} \rhounit$ for galaxies with stellar mass in the range $10^{10}-10^{11} \Msun$ at $z=0$ \citep[see Figure 3 of ][]{Motwani:2022}.    
Note that the SFR-weighted mean value $\Sigma_*\sim 10 \surfunit$ in $z=0$ galaxies is lower than that observed in nearby star-forming galaxies; in the PHANGS survey, for example, this is closer to $\Sigma_*\sim 100 \surfunit$ \citep{2022AJ....164...43S}. While TNG rotation curves are not dissimilar to those in observed $z=0$ disk galaxies, \citet{2018MNRAS.481.1950L} previously pointed out that the dark matter contribution to the potential in TNG galaxies at small radii appears to exceed that of observed disk galaxies, \REV{which is reflected in the peak of the $\Sigma_*$ distribution appearing at a significantly lower  value than in nearby galaxies.  Potentially, more realistic models of star formation and galactic winds may rectify this discrepancy}.

The basic properties shown here are those that are needed as input to implement the PRFM model in TNG50. We next discuss a key element of the PRFM theory, namely the vertical equilibrium requirement.

\section{Vertical Equilibrium and Dynamical Timescale in the ISM of Disk Galaxies}\label{sec:equil}

In this section, we derive relations for the equilibrium midplane pressure, vertical thickness, and vertical dynamical time for a gas disk subject to a gravitational potential with contributions from the gas, the stellar disk, and a spherical dark matter halo.  These relations generalize those presented in Section 2.1 of \citetalias{2022ApJ...936..137O}, and represent a time-averaged, locally horizontally-averaged state of the ISM disk, such that the relevant variables are treated just as functions of vertical coordinate $z$. 

In \citetalias{2022ApJ...936..137O}, which focused on conditions as found in local star-forming galaxies, the results presented were for the case in which the gas disk is thin compared to the stellar disk.  In that circumstance, the gravitational effects of the stellar disk and the dark matter potential can be captured using their combined  midplane density.  Since, however, the assumption of a thin gas disk may not hold in high-redshift starburst galaxies (as well as low-redshift analogs), here we derive more general formulae than those presented in \citetalias{2022ApJ...936..137O}. In particular, the analysis here allows for the gas disk to be either thinner or thicker than the stellar disk, and also allows for arbitrary relative importance of the gravity from gas, stars, and dark matter in confining the gas.   \REV{To make contact with other formulae that have previously appeared in the literature (and for the convenience of readers who prefer a simpler expressions where applicable), in the solutions presented here we include formulae for limiting cases as well as the most general case.}  

\REV{In numerical simulations of galaxy formation, the vertical thickness of the disk is sometimes well resolved and sometimes unresolved.  \autoref{sec:res_usage} discusses resolution criteria, and provides a guide to results that may be used in resolved/unresolved cases.}

\subsection{\REV{ISM Pressure and Weight}}

We can write the effective pressure that provides vertical support in the gas disk, $\Ptot$, as the  product of the density, $\rhog$, and square of the effective velocity dispersion $\seff$: $\Ptot = \rhog \seff^2$.  
Here, $\Ptot$ (and therefore  
$\seff$) may include thermal, turbulent, and magnetic contributions, defined using (horizontal) averages as 
$\Ptot = \Pth + \Pturb + \Pimag$ for 
$\Pth\equiv \langle \rhog c_s^2\rangle $ (thermal pressure),
$\Pturb \equiv \langle \rhog v_z^2 \rangle$ (turbulent pressure),
and $\Pimag \equiv \langle|{\bf B}|^2 - 2B_z^2\rangle/(8\pi) =
\langle B_x^2 + B_y^2 - B_z^2 \rangle/(8\pi)$ (vertical Maxwell stress,
combining magnetic pressure and tension).

In equilibrium, the midplane pressure must be equal to the vertical weight of the ISM,
\begin{equation}
\Ptot = \mathcal{W} \equiv \int_{0}^{z_{\rm max}}\rhog(z)g(z)dz
\end{equation}
for $g(z)=\partial{\Phi}/\partial{z}$ the total vertical gravity, with contributions from gas, stars, and dark matter, $g=g_\mathrm{g} + g_\mathrm{*} + g_\mathrm{d}$.   Here and elsewhere, we abbreviate $\Ptot(z=0)\rightarrow \Ptot$. The agreement between $\Ptot$ and $\mathcal{W}$ has been well documented in many simulations \citep[e.g.,][]{2007ApJ...663..183P,2013ApJ...776....1K,2015ApJ...815...67K,2016MNRAS.462.3053B,2020ApJ...894...12V,2020ApJ...898...35K,2020MNRAS.498.3664G,2022ApJ...936..137O}.

\REV{For any disk, the total gas surface density,  
$\Sg$, is considered a known quantity. Since this is a vertical integral of the density, it is not subject to any theoretical assumptions regarding the shape of the vertical profile, and in numerical simulations can be computed robustly independent of resolution by projection perpendicular to the midplane.  
The effective half-thickness of the disk, $\Hg$, is defined from $\Sg$ and the midplane density, $\rhog(0)$, such that }
$\rhog(0) = \Sg/(2 \Hg)$.  
In equilibrium we therefore have midplane pressure
\begin{equation}\label{eq:pressure-weight}
  \Ptot = \frac{\Sg}{2 \Hg} \sigma_{\rm eff}^2 
  = {\cal W}_\mathrm{g} + {\cal W}_* + {\cal W}_\mathrm{d}.
\end{equation}

The contribution to the gas weight from the gas gravity is 
\begin{eqnarray}\label{eq:gas_W}
    {\cal W}_\mathrm{g} &=& \int_0^{z_{\rm max}}\rhog g_\mathrm{g} dz = \frac{1}{4\pi G} \int_0^{z_{\rm max}}\pderiv{g_\mathrm{g}}{z}g_\mathrm{g} dz\nonumber \\
    &=& \frac{1}{8\pi G} g_\mathrm{g}( z_{\rm max})^2 = \frac{\pi G \Sg^2}{2}.
\end{eqnarray}
We have assumed plane-parallel geometry, in which $g(z_{\rm max}) = 2 \pi G \Sg$.

The contribution to the gas weight from stellar disk's gravity is
\begin{equation}
    {\cal W}_* = \int_0^{z_{\rm max}}\rhog g_* dz = \frac{1}{4\pi G} \int_0^{z_{\rm max}}\pderiv{g_\mathrm{g}}{z}g_* dz.
\end{equation}
If we define $\tilde g \equiv g(z)/g(z_{\rm max}) = g(z)/(2\pi G \Sigma)$ for either stars or gas, this becomes
\begin{equation}\label{eq:Wstar_general}
    {\cal W}_* =  \pi G \Sg \Sigma_* \int_0^{z_{\rm max}}\pderiv{\tilde{g}_\mathrm{g}}{z}\tilde{g}_* dz.
\end{equation}
The  value of the integral using the normalized gravitational profile functions depends on their detailed shape; it is equal to 1/2 when the profiles are the same. Since $max(\tilde{g}_*)=1=\tilde{g}_\mathrm{g}(z_\mathrm{max})$ and $\tilde{g}_\mathrm{g}(0)=0$, the integral is bounded above by unity.  A good approximation is given by 
\begin{equation}\label{eq:stellar_W}
    {\cal W}_* \approx \pi G \Sg \Sigma_* \frac{\Hg}{\Hg + H_*},  
\end{equation}
where $H_*\equiv \Sigma_*/[2 \rho_*(0)]$, analogous to the definition of $H_\mathrm{g}$ above.\footnote{We note that this $H_*$ is {\it defined} in terms of the stellar total surface density and midplane volume density; care must be taken as this convention may differ from conventions adopted for the functional forms empirically fit to the vertical distribution of stars \citep[e.g.][]{1988A&A...192..117V}.}
\autoref{eq:stellar_W} is exact in the case that the vertical profiles are exponential. For Gaussian gas and stellar disk density profiles, 
$H_\mathrm{g}/(H_\mathrm{g}+ H_*)$ would be replaced by $(2/\pi) \tan^{-1}(H_\mathrm{g}/H_*)$, yielding a result at most 25\% less than or 6\% greater than that in \autoref{eq:stellar_W} for $H_\mathrm{g}/H_* > 0.2$.  

For a spherical dark matter distribution\footnote{The contribution to the ISM weight due to the gravity of a spherical stellar bulge takes the same form as that due to a dark matter halo, i.e. $\mathcal{W}_\mathrm{b} =\zeta \Sg \Omega_\mathrm{b}^2 \Hg$ for $\Omega_\mathrm{b}^2 = r^{-1} \partial\Phi_\mathrm{b}/\partial r $, with $\Phi_\mathrm{b}$ the bulge potential.  For a uniform-density bulge, $\Omega_\mathrm{b}^2=4\pi G \rho_\mathrm{b}/3$, and for a bulge with a Hernquist profile, $\Omega_\mathrm{b}^2=2 \pi(1+r/r_\mathrm{b}) G \rho_\mathrm{b}(r)$.}, 
\begin{equation}\label{eq:dark_W}
 {\cal W}_\mathrm{d} =  \int_0^{z_{\rm max}}  \rhog \Omega_\mathrm{d}^2 z dz   = \zeta \Sg \Omega_\mathrm{d}^2 \Hg,
\end{equation}
where we have assumed $\Hg\ll r$; here $\Omega_\mathrm{d}\equiv V_\mathrm{c}/r$ is the angular rotation velocity associated with the dark matter (i.e. $\Omega_\mathrm{d}^2 (r)= r^{-1}\partial\Phi_\mathrm{d}/\partial r$), and  \REV{$\zeta \approx 1/3$.  This value of $\zeta$  applies for gas disks that are confined either primarily by external gravity or primarily by self-gravity, provided $\Hg\ll r$} \citep{2011ApJ...731...41O}.  
For a flat rotation curve, $\Omega_\mathrm{d}^2 = 4 \pi G \rho_\mathrm{d}$ may be used, for $\rho_\mathrm{d}$ the local dark matter density.

\REV{
Inserting \autoref{eq:gas_W}, \autoref{eq:stellar_W}, and \autoref{eq:dark_W} in  \autoref{eq:pressure-weight}, we obtain 
\begin{equation}\label{eq:combined_Hg_eq}
  \Hg\left[ 1 + \frac{\Sigma_*}{\Sg} \frac{2\Hg}{\Hg+H_*} + \frac{2 \zeta \Omega_\mathrm{d}^2 }{\pi G \Sg} \Hg\right] = \frac{\seff^2}{\pi G \Sg}. 
\end{equation}
In the next two subsections, we provide solutions of this equation for $\Hg$ in various limits, as well as the general solution.
}

In the formulae for $\Hg$ presented in \autoref{sec:H_W_solns}, it is assumed the stellar disk thickness $H_*$ is known, either through direct measurement (in resolved simulations or observations) or through an empirical relationship such as a fixed ratio between vertical and radial scale length.  For the latter, based on nearby-universe observations \citep[see e.g.][]{1982A&A...110...61V,2002MNRAS.334..646K,2020ApJ...892..148S}, the most commonly adopted choice corresponds to $H_*=0.27 l_*$ for $l_*$ the exponential radial scale length. 

\REV{In \autoref{sec:equal_thickness}, we provide results for an alternative situation in which $H_*$ is not 
directly known, but may be assumed to be comparable to $\Hg$.}  

\subsection{\REV{Solutions for $\Hg$ and $\cal W$}} \label{sec:H_W_solns}

If we consider just the gas and stellar disk terms in the weight, which typically dominate within the star-forming disks of observed galaxies at low redshift, 
the third term in the square brackets of \autoref{eq:combined_Hg_eq} may be dropped.  

The solution of the  resulting quadratic equation is
\begin{eqnarray}\label{eq:Hg_general}
    \Hg^{\rm gas+star} = \hskip2.5in \nonumber\\ 
    \frac{2 \sigma_{\rm eff}^2 }{\pi G  \Sg - \frac{\sigma_{\rm eff}^2}{H_*} + 
    \left[\left(\pi G  \Sg + \frac{\sigma_{\rm eff}^2}{H_*}\right)^2 + 8\pi G\Sigma_* \frac{\sigma_{\rm eff}^2}{H_*}  \right]^{1/2}},\hskip0.25in
\end{eqnarray}
where the superscript indicates that only the potential of gas and stars is taken into account.
Note that in the gas-only limit, this recovers the familiar result $H_{\rm g}^{\rm gas-only} = \sigma_{\rm eff}^2/(\pi G \Sigma_g)$.  In the limit where only stellar gravity is considered, we obtain $H_\mathrm{g}^{\rm stellar-grav}=\left[1+ \left( 1+ 8 \pi G \Sigma_* H_*/\seff^2\right)^{1/2}  \right]\seff^2/(4\pi G \Sigma_*)$; in the thin gas disk limit $H_\mathrm{g}/H_*\ll 1$ this becomes $H_\mathrm{g}^{\rm stellar-grav} \rightarrow \sigma_{\rm eff}/(4 \pi G \rho_*)^{1/2}$, while in the 
thick gas disk limit $H_\mathrm{g}/H_*\gg 1$ this becomes 
$H_\mathrm{g}^{\rm stellar-grav} \rightarrow \sigma_{\rm eff}^2/(2\pi G \Sigma_*)$.

\autoref{eq:Hg_general} may be substituted back for $\Hg$ in \autoref{eq:stellar_W} to obtain the weight in the stellar potential  in terms 
of the gas parameters $\Sg$ and $\seff$, and the 
stellar disk parameters $\Sigma_*$ and $H_*$.  
The result is 
\begin{eqnarray}
\label{eq:Wstar}
{\cal W}_*^{\rm gas+star} = \hskip2.5in \nonumber\\ 
\frac{2 \pi G \Sg \Sigma_* }{1 + \frac{\pi G \Sg H_*}{\seff^2} + \left[  \left(1+ \frac{\pi G \Sg H_*}{\seff^2}\right)^2 + \frac{8 \pi G \Sigma_* H_*}{\seff^2} \right]^{1/2}}\hskip0.25in
\end{eqnarray}
A convenient approximate expression for the total weight (sum of gas and stellar terms) in the absence of a dark matter contribution is 
\begin{equation}
\label{eq:approx_weight}
{\cal W}^\mathrm{gas+star} 
\approx  \frac{\pi G \Sg^2}{2} + \frac{2 \pi G \Sg \Sigma_* }{1 + \left[1+ \frac{8 \pi G \Sigma_* H_*}{\seff^2}  \right]^{1/2}};
\end{equation}
we note that the terms dropped from the denominator of \autoref{eq:Wstar} in reaching \autoref{eq:approx_weight} affect the value of ${\cal W}_*$ only when it is subdominant compared to ${\cal W}_\mathrm{g}$.  
In the limit where $\Hg \ll H_*$ (for small $\seff$),  the second term in \autoref{eq:approx_weight} becomes $\Sg(\pi G \rho_*)^{1/2}\seff$; this is slightly larger than the commonly adopted expression ${\cal W}_*\approx \Sg(2 G \rho_*)^{1/2}\seff$ \citepalias[e.g. Equation 7 of][]{2022ApJ...936..137O}, which assumes a Gaussian vertical gas profile. When $\seff$ is large, however, as may occur in starbursting regions which have $\Hg/H_* >1$ \citep[see e.g.][]{2021ApJ...909...12G}, the conventional form $\Sg(2 G \rho_*)^{1/2}\seff$ would significantly overestimate ${\cal W}_* $, which from \autoref{eq:Wstar_general} has an upper limit $\pi G \Sg \Sigma_*$. The expression in \autoref{eq:approx_weight} automatically imposes this constraint and captures both limits of $\Hg/H_*$ mentioned above.  

In the case that the stellar disk is negligible and there is only a gas disk and dark matter halo \REV{(which approximates the situation of some very low surface brightness, gas-rich dwarfs)},
we drop the second term in square brackets in \autoref{eq:combined_Hg_eq},
and the resulting quadratic has solution
\begin{equation}\label{eq:Hg_gas_dm}
    \Hg^{\rm gas+dm} = 
    \frac{2 \sigma_{\rm eff}^2 }{\pi G  \Sg  + 
    \left[\left(\pi G  \Sg \right)^2 + 8\zeta\Omega_\mathrm{d}^2 \sigma_{\rm eff}^2  \right]^{1/2}};\hskip0.25in
\end{equation}
note that we may substitute $\Omega_\mathrm{d} \rightarrow \Omega_\mathrm{b}$ in this expression for the case of a gas disk plus stellar bulge.

\REV{In the most general case, the terms for gaseous, stellar, and dark matter gravity in \autoref{eq:combined_Hg_eq} are all retained, and}
one must solve a cubic for $\Hg$:
\begin{eqnarray}\label{eq:Hg_cubic}
    \Hg^3 \left(  \frac{2\zeta\Omega_\mathrm{d}^2  }{\pi G \Sigma_g} \right) +
    \Hg^2 \left(1 + \frac{2\Sigma_*}{\Sg} + \frac{2\zeta\Omega_\mathrm{d}^2  H_*}{\pi G \Sg} \right)\nonumber\\  
    + \Hg \left(H_* - \frac{\sigma_{\rm eff}^2 }{\pi G\Sg} \right) - H_* \frac{\sigma_{\rm eff}^2 }{\pi G\Sg}=0. 
\end{eqnarray}
To obtain the solution, we first divide out by the coefficient of the leading term (which has units $L^{-1}$) so that the coefficients of $\Hg^2$, $\Hg^1$, and $\Hg^0$ now read:
\begin{equation}
  a_{2} = \left(1+ \frac{2\Sigma_*}{\Sg} + \frac{2\zeta \Omega_\mathrm{d}^2 H_*}{\pi G \Sg} \right) \left(  \frac{2\zeta \Omega_\mathrm{d}^2 }{\pi G \Sg} \right)^{-1} 
\end{equation}
\begin{equation}
  a_{1}  = \left(H_* - \frac{\sigma_{\rm eff}^2 }{\pi G\Sg} \right)  \left(  \frac{2\zeta \Omega_\mathrm{d}^2 }{\pi G \Sg} \right)^{-1} 
\end{equation}
\begin{equation}
  a_{0}  =  - H_* \frac{\sigma_{\rm eff}^2 }{\pi G\Sg} \left(  \frac{2\zeta \Omega_\mathrm{d}^2 }{\pi G \Sg}\right)^{-1}. 
\end{equation}
We then define:
\begin{equation}
    Q = \frac{3 a_{1} - a^{2}_{2}}{9},\,\,  R = \frac{9 a_{2} a_{1} - 27 a_{0} - 2a_{2}^{3}}{54}
\end{equation}
The discriminant, $D = Q^{3} + R^{2}$, in this case is less than zero, which means there are three  different {\it real} solutions. However, only one solution is positive:
\begin{equation}\label{eq:Hg_withdm}
    \Hg = 2 \sqrt{-Q} \cos\left( \frac{\theta}{3}   \right) - \frac{1}{3} a_{2},\,\, \theta =  \cos^{-1}\left(\frac{R}{\sqrt{-Q^{3}}}\right).
\end{equation}

We note that  the thickness of the gas disk can alternatively be expressed in terms of the gas and stellar volume densities, rather than surface densities (as in \autoref{eq:Hg_cubic}).  This may be obtained using 
\begin{equation}\label{eq:Hrho}
\Hg= \frac{\seff}{\left[2\pi G \rhog + \frac{4 \pi G \rho_*}{1+\Hg/H_*} + 2 \zeta \Omega_\mathrm{d}^2\right]^{1/2}},    
\end{equation}
which may be solved iteratively (or through direct solution of the corresponding cubic), given a value of $H_*$.
The dark matter term in the denominator is $(8\pi/3)\zeta G \rho_\mathrm{d}$ for a flat rotation curve. 
In the typical case where  $\Hg/H_* \lesssim 1$, \autoref{eq:Hrho} implies $\Hg$ varies inversely as the square root of a weighted sum of gas, stellar, and dark matter densities.

Once $\Hg$ is obtained from  \autoref{eq:Hg_withdm}, 
the equilibrium pressure is given by $\Ptot= {\cal W} = \seff^2 \Sg/(2\Hg)$ \REV{as in \autoref{eq:pressure-weight}.  The above assumes that $\seff$ is given.  If, instead, $\seff$ is a function of $\Ptot$, the above procedure is iterated to find self-consistent equilibrium values of $\seff$, $\Hg$, and ${\cal W}=\Ptot$.}

\subsection{\REV{{$H$} and {$\cal W$} for Equal-Thickness Disks}}\label{sec:equal_thickness}

In cosmological simulations of galaxies,  a proper measure of $H_*$ may be not be available due to lack of resolution.  In this circumstance, an alternative to the solution in  \autoref{eq:Hg_withdm} is needed to obtain $\Hg$ from direct measurements in the simulation, since the coefficients in \autoref{eq:Hg_cubic} require a value for $H_*$.  Given that stars form out of gas and relatively little kinetic heating of the stellar distribution has occurred at early epochs,  a reasonable zeroeth-order assumption\footnote{Alternative closure assumptions may be adopted; for example, one might adopt a relation between $\seff$ and the stellar velocity dispersion $\sigma_*$  -- see e.g. \citet{2023arXiv230207823F} -- and use $H_*/H_\mathrm{g} = (\sigma_*/\seff)^2$.} is that $H_* \sim \Hg$. In the special case $H_*=\Hg=H$, the cubic of \autoref{eq:Hg_cubic} reduces to a quadratic, with solution 
\begin{eqnarray}\label{eq:H_special}
H = \frac{2\seff^2}{\pi G(\Sg + \Sigma_*)} \times \hskip1.2in \nonumber \\ 
\left(1 
+ \left[
1 +
 \frac{8\zeta \Omega_\mathrm{d}^2\seff^2}{(\pi G)^2(\Sg + \Sigma_*)^2}\right]^{1/2}\right)^{-1}.
\end{eqnarray}
We note that this special case is the same as \autoref{eq:Hg_gas_dm} with $\Sg \rightarrow \Sg + \Sigma_*$; a bulge term could also be included by substituting $\Omega_\mathrm{d}^2 \rightarrow \Omega_\mathrm{d}^2 + \Omega_\mathrm{b}^2$.  

For the special case where the stellar and gas disks are assumed to have the same thickness, the result for ${\cal W}=\Ptot = \seff^2 \Sg/(2H)$ is
\begin{eqnarray}
\label{eq:H_comb_disk}
\mathcal{W} = \frac{\pi G\Sg(\Sg + \Sigma_*)}{4} \times \hskip1.2in \nonumber \\ 
\left(1 
+ \left[
1 +
 \frac{8\zeta \Omega_\mathrm{d}^2\seff^2}{(\pi G)^2(\Sg + \Sigma_*)^2}\right]^{1/2}\right).
\end{eqnarray}
\REV{As above, this assumes $\seff$ is given. If instead $\seff$ is a known function of $\Ptot$, \autoref{eq:H_comb_disk} and the $\seff-\Ptot$ relation would be iterated to reach a solution.}


\subsection{\REV{Dynamical Timescale}}\label{sec:tdyn}

We shall define the (vertical) dynamical time as 
\begin{subequations}
\begin{eqnarray}\label{eq:tverta}
    t_{\rm dyn} &\equiv& \frac{2 \Hg }{\sigma_{\rm eff}}\\
    \label{eq:tvertb}
    &=& \frac{\Sigma_g \sigma_{\rm eff}}{\Ptot} = \frac{\Sg \seff}{{\cal W}_\mathrm{g} + {\cal W}_* + {\cal W}_\mathrm{d}}.
\end{eqnarray}
\end{subequations} 
In the case that the stellar disk thickness $H_*$ is known but the gas  
disk thickness is uncertain (in observations) 
or unresolved (in simulations), 
$\Hg$ as given from  \autoref{eq:Hg_withdm} 
should be used in \autoref{eq:tverta}.  If dark matter is unimportant to the vertical gravity, \autoref{eq:Hg_general} could be used in place of  \autoref{eq:Hg_withdm}. 
If $H_*$ is also uncertain or unresolved, \autoref{eq:H_special} could instead be used for $\Hg$, provided it is reasonable to assume $H_*\sim \Hg$.  The theoretical expressions for $\Hg$ (i.e. \autoref{eq:Hg_general}, \autoref{eq:Hg_withdm}, \autoref{eq:H_special}) employ the total ISM gas and stellar surface densities $\Sg$ and $\Sigma_*$, which can be robustly measured in a simulation even if the resolution is low.

An alternative expression for the vertical dynamical time, obtained by using \autoref{eq:Hrho} 
in \autoref{eq:tverta}, is:
\begin{equation}\label{eq:tdyn_rho}
t_\mathrm{dyn} = \frac{2}{\left[2 \pi G \rhog 
+ \frac{4 \pi G \rho_*}{1 + \Hg/H_* } + 
2 \zeta \Omega_\mathrm{d}^2
\right]^{1/2}},   
\end{equation}
where \autoref{eq:Hrho} can be used to obtain $\Hg/H_*$ if $H_*$ is known, and $2 \Omega_\mathrm{d}^2 \rightarrow (8\pi/3)G \rho_\mathrm{d}$ for a flat rotation curve.    
If a bulge is significant, it may be included by replacing 
$ \Omega_\mathrm{d}^2 \rightarrow \Omega_\mathrm{d}^2 + \Omega_\mathrm{b}^2 $.  
For the special case where $H_*=\Hg$, the term involving the stellar density becomes $2\pi G \rho_*$.

\autoref{eq:tdyn_rho} shows that the dynamical time in general depends on a weighted sum of the gas, stellar, and dark matter densities, which appear on essentially an equal footing.  
In nearby normal spiral galaxies, the largest term 
is often that involving the stellar density, but this is 
not necessarily the case in high redshift 
galaxies (for which the gas density may 
dominate), or in low surface brightness dwarfs (for which the dark matter 
density term may dominate).  

It is important to recognize that \autoref{eq:tdyn_rho} will overestimate the true dynamical time if the densities are lower than they should realistically be.  This would be the case, for example, in simulations where the physical resolution is too low compared to what $\Hg$ should be (as predicted from \autoref{eq:Hg_general} or \autoref{eq:Hg_withdm} or \autoref{eq:H_special}), so that numerical diffusion and/or gravitational softening thicken the disk and reduce $\rhog$ below what it should be for a given $\Sg$.  Thus, \autoref{eq:tdyn_rho} can only be used if the 
true vertical thickness of the disk is resolved, which for simulations means being converged with respect to decreases in the physical scale of the numerical grid or the adopted mass resolution. 

\subsection{\REV{Resolution Requirements in Simulations and a Guide to Usage }
}\label{sec:res_usage}

To provide some idea of the numerical resolution that would be needed in order to use \autoref{eq:tdyn_rho}, we consider \autoref{eq:H_special} for a disk of stars and gas (as appropriate for low-redshift galaxies, in which the dark matter term is typically only $\sim 10\%$ of the stellar plus gas term, \REV{and $\Hg\sim H_*$}), 
\begin{equation}\label{eq:H_dimen}
 H=110 \pc \left(\frac{\seff}{15\kms}\right)^2  \left(\frac{\Sigma_* + \Sg}{150 \surfunit}\right)^{-1}. 
\end{equation}
The fiducial surface density value in the above is motivated by the resolved properties of PHANGS star-forming galaxies in the local Universe \citep{2022AJ....164...43S}, which have mean $\Sigma_*=110 \surfunit$, $\Sg=28\surfunit$ (when weighted by molecular gas mass, which is similar to weighting by star formation).\footnote{When weighted by area, the mean values within the PHANGS sample are instead $\Sigma_*=65 \surfunit$ and $\Sg=13\surfunit$, implying a factor of two larger $H$ than obtained with the fiducial parameters of  \autoref{eq:H_dimen}.}    The fiducial velocity dispersion is motivated by observations of CO and \ion{H}{1} velocity dispersions in resolved nearby galaxies \citep[see][and references/discussion in \autoref{sec:prfm}]{2016AJ....151...15M}, which when mass-weighted in quadrature yield $\sigma=9\kms$; this is likely enhanced by another $\sim 50\%$ when magnetic terms are included \citep{2022ApJ...936..137O,2023ApJ...946....3K}.  Using the PHANGS numbers for surface densities and $\seff=15\kms$, 
to (marginally) resolve the full disk thickness ($2H$) vertically by 4 elements would require a cell (or particle) mass of $\Sg H^2/16 \rightarrow 2.5 \times 10^4 \Msun$; this would increase to $\sim 10^5 \Msun$ for conditions similar to the solar neighborhood, in which $\Sg$ is $\sim 3$ times smaller and $H$ is $\sim 3$ times larger.\footnote{By comparison, the mean baryon mass resolution is $8.5 \times 10^4 M_\odot$, $1.4 \times 10^6 M_\odot$, and $1.1 \times 10^7 M_\odot$  
in TNG50, TNG100, and TNG300, respectively -- see  {\tt https://www.tng-project.org/about/}.}

More generally, in order to resolve the disk thickness vertically by $N_d$ cubic cells each of side length  $L=2H/N_d$, the mass in each cell would need to be 
\begin{eqnarray}\label{eq:massres}
m_{\rm cell}&=&4 H^2 \Sg/N_d^3 \nonumber\\
&=& 
\frac{4}{N_d^{3}}\frac{\seff^4 \Sg }{(\pi G)^{2}(\Sg + \Sigma_*)^2}.  
\end{eqnarray}
Including a dark matter contribution \REV{to the gravity confining the disk vertically} would reduce $H$, making the mass resolution requirement more stringent. \REV{
For very high surface density conditions, as prevail at high redshift and are also present in starburst regions at low redshift, the velocity dispersion will also generally be higher.  Whether the requirement for the disk to be resolved becomes more or less stringent depends on the eEoS that is adopted. In the case of a power-law barotropic eEoS, for which $\seff\equiv(\Ptot/\rho)^{1/2}\propto \Ptot^\beta$, the scaling $\Ptot = {\cal W} \propto \Sigma^2$ implies that the minimum $m_\mathrm{cell}$ would tend to increase at higher pressure provided $\beta >0.125$, corresponding to a pressure vs. density scaling stiffer than $4/3$.}

\REV{In \autoref{sec:prfm}, we describe a theoretical characterization of star formation 
that depends on the dynamical time $\tdyn$ and a coefficient whose factors have been calibrated as a function of pressure 
$\Ptot$ in high-resolution ISM simulations (see \autoref{eq:SFR_PRFM}).  To use this as a prescription for star formation in a cosmological galaxy formation simulation, it is necessary to have  local measures of $\Ptot$ and $\tdyn$.  How these are estimated from quantities 
that are available in the simulation depends on whether the galactic ISM disk is vertically resolved or not.  We can distinguish three application cases, as follows:

\begin{itemize}

\item[(1)] In the case that {\it both the ISM disk and the stellar disk are vertically resolved}, projections perpendicular to the local disk plane would first be needed to compute surface densities $\Sg$ and $\Sigma_*$.  The half-thicknesses would be set to $\Hg=\Sg/(2\rhog)$ and $H_*=\Sigma_*/(2\rho_*)$, where $\rhog, \rho_*$ are measured midplane densities; $\Hg$ should agree with \autoref{eq:Hrho}.  Then, \autoref{eq:tdyn_rho} would be employed for $\tdyn$, using the measured densities and $\Omega_\mathrm{d}$. 
The pressure would be set to $\Ptot=\seff^2 \rhog$.    

\item[(2)] In the case that {\it the ISM disk is unresolved but the stellar disk is resolved}, projections perpendicular to the local disk plane would first be needed to compute surface densities $\Sg$ and $\Sigma_*$, and the stellar half-thickness would be set to $H_*=\Sigma_*/(2\rho_*)$ where $\rho_*$ is the local measured stellar density.  Then, \autoref{eq:Hg_withdm} would be used for $\Hg$, and \autoref{eq:tverta} would be used for $\tdyn$.  The pressure would be set to $\Ptot = \seff^2\Sg/(2\Hg)$.

\item[(3)] In the case that {\it both the ISM disk and the stellar disk are vertically unresolved}, projections perpendicular to the local 
disk plane would first be needed to compute $\Sg$ and $\Sigma_*$.  Then, under the assumption that $\Hg\approx H_*$ is satisfactory, \autoref{eq:tverta} would be used for $\tdyn$, with \autoref{eq:H_special} for $\Hg=H$.  The pressure would be set to $\Ptot =\seff^2\Sg/(2H)$, i.e. to the value in  \autoref{eq:H_comb_disk}.  
\end{itemize}

If our prescription (see \autoref{sec:prfm}) for the calibration of $\seff$ as a function of $\Ptot$ is also adopted as an eEoS and $\Hg$ is resolved, this can be used to set the pressure in the simulation, given 
$\rhog$. In cases (2) and (3) of vertically unresolved gas disks, 
the equilibrium estimate $\rhog= \Sg/(2\Hg)$ must be used with \autoref{eq:Hg_withdm} or \autoref{eq:H_special} for $\Hg$ (rather than the measured $\rhog$, which would be an underestimate), and iteration is required since $\Hg$ depends on $\seff$. 
}

\section{Summary of PRFM Theory and Subgrid Model Calibration from TIGRESS Simulations}\label{sec:prfm}

\subsection{The Depletion Time}\label{sec:tdep}

In the PRFM theory, gas pressure in a disk responds to star formation feedback as 
\begin{equation}\label{eq:yield_def}
\Ptot = \Upstot \SSFR,    
\end{equation}
where the feedback yield $\Upstot$ (which has units of velocity) includes terms from thermal pressure (arising from radiation heating), kinetic turbulent pressure (arising from supernova blast waves), and magnetic pressure (responding 
to the kinetic turbulence).  
\REV{Physically, \autoref{eq:yield_def} represents a balance between energy gains and losses of various forms in the ISM \citepalias[see Section 2 of][and references therein]{2022ApJ...936..137O}.  For example, equilibrium between radiative heating  and cooling would lead to $P_{\rm th}= \Gamma kT/\Lambda$ where $\Gamma$  is the heating rate coefficient and $\Lambda$ is the cooling rate coefficient. Since radiative heating is proportional to the UV radiation field strength produced by young stars,  we have $\Gamma \propto \SSFR$, leading to $P_{\rm th}=\Upsilon_{\rm th} \SSFR$}.\footnote{\REV{Here, the coefficient $\Upsth$ absorbs the functional dependence of $\Gamma/\Lambda$ on the UV luminosity-to-young star mass ratio, radiative transfer subject to metal and dust abundances, and radiation/gas interaction crossections and cooling rate coefficients affected by detailed ISM properties -- see \citet{2023ApJS..264...10K,2024arXiv240519227K}}.}

The depletion time averaged over all of the gas is then 
\begin{equation}\label{eq:tdep_PRFM}
 t_\mathrm{dep} =  \frac{\Sg}{\SSFR} = \Upstot \frac{\Sg}{\Ptot}
 = \frac{\Upstot}{\seff} t_\mathrm{dyn},
\end{equation}
where we use $\Sg = 2 \Hg \rhog$, $\Ptot =\seff^2 \rhog$, and $t_\mathrm{dyn} \equiv 2\Hg/\seff$ (see \autoref{eq:tverta}).  Provided that vertical dynamical equilibrium is satisfied, the pressure will be equal to the weight of the ISM, and the thickness of the disk will be consistent with its equilibrium prediction.  
The mean gas depletion time in equilibrium can then be expressed in terms of mean values of the feedback yield, effective velocity dispersion, and vertical dynamical time.  

For cosmological simulations, the above provides a prediction for the SFR $\dot{m}_*$ in a cell that averages over the (spatially and temporally unresolved) lifecycle of star formation and feedback energy return in multiphase gas:
\begin{equation}\label{eq:SFR_PRFM}
 \dot{m}_* =\frac{m_\mathrm{g}}{\tdep}= \frac{\seff}{\Upstot}\frac{m_\mathrm{g}}{\tdyn},\end{equation}
where $m_\mathrm{g}$ is the gas mass in an individual cell (or particle).  
For practical use as a star formation prescription, it is necessary to have calibrated predictions for $\seff$ and $\Upstot$ as a function of parameters that can be robustly measured, even at low resolution, in the cosmological simulation, \REV{as discussed below}.  

\REV{Practical use of \autoref{eq:SFR_PRFM} also requires} a measure of $\tdyn$.
Calculation of $\tdyn$ in equilibrium is discussed in \autoref{sec:tdyn}. As noted there, in general  $\tdyn$ depends not just on the gas 
density, but also on the stellar density and dark matter density and some measure of the relative thickness of the gas and stellar disks. 
When
the stellar and gas disks are at least marginally resolved in the cosmological simulation (which would typically require baryon mass resolution $\sim 10^4-10^5\Msun$; see \autoref{eq:H_dimen} and subsequent text), $\tdyn$ may be computed directly using \autoref{eq:tdyn_rho} from the simulation variables $\rhog$, $\rho_*$, and $\rho_\mathrm{d}$, along with direct measurements of $H_*$ and $\Hg$ (e.g. from local vertical gradients of $\rho_*$ and $\rhog$). \REV{This is case (1) in \autoref{sec:res_usage}.} 

At coarser mass resolution ($\gtrsim 10^6\Msun$), the densities and pressure measured in the simulation would underestimate the true values in a real galaxy with the same macroscopic properties, and  using measured values of $\rhog$ and $\rho_*$ from the simulation in \autoref{eq:tdyn_rho} (or \autoref{eq:Hrho}) would result in an overestimate of what the true dynamical time (or gas disk thickness) should be.  In this situation, \autoref{eq:tdyn_rho} 
cannot be used for $\tdyn$.   However, given measures of $\Sg$ and $\Sigma_*$ (obtained by integrating through the disk, which in cosmological simulations requires identifying the direction normal to the disk plane), one may use \autoref{eq:Hg_withdm} (if $H_*$ can be estimated) in order to {\it predict} the ISM disk thickness $\Hg$ and then $\tdyn$ from \autoref{eq:tverta}; \REV{ this is case (2) in \autoref{sec:res_usage}.} If it is not possible to obtain a direct estimate of $H_*$, one may instead use    \autoref{eq:H_special} for the {\it predicted} disk thickness $H=\Hg=H_*$; \REV{ this is case (3) in \autoref{sec:res_usage}.}
In general, a {\it predicted} value of $\seff$ from a subgrid ISM eEoS model is also needed.  

\REV{In the situation when dark matter is unimportant to vertical disk confinement and $\Hg\approx H_*$, the depletion time and pressure, $\tdep$ and  $\Ptot={\cal W}$, have particularly simple forms:
\begin{equation}\label{eq:simple_tdep}
\tdep= \Upstot \frac{2}{\pi G (\Sg + \Sigma_*)},\ \  
\Ptot=\frac{\pi G \Sg (\Sg + \Sigma_*)}{2}.    
\end{equation}
}
\REV{The effective velocity $\seff$ dispersion does not enter either expression.  Because $\Upstot$ has been calibrated in terms of $\Ptot$ (see below), the only quantities that are required in order to obtain the predicted gas depletion time in this case are the stellar and gas surface densities.}

\subsection{Evaluation of  $\Upstot$ and $\seff$}\label{sec:calibrations}

Both the feedback yield $\Upstot$ and the effective velocity dispersion $\seff$ are based on averages over multiple ISM phases and components, and respond to a complex array of physical effects. While a rough estimate of $\Upstot$ may be obtained from simple theoretical considerations  \citep[see][for analytic estimates of the thermal and turbulent yield, respectively]{2010ApJ...721..975O,2011ApJ...731...41O}, more accurate values for both quantities, and their dependence on galactic environment, require calibration from high-resolution ISM simulations with realistic modeling of the multiphase ISM, star formation, and feedback \citep[see][and below]{2011ApJ...743...25K,2013ApJ...776....1K,2015ApJ...815...67K,2017ApJ...846..133K,2020ApJ...900...61K,2023ApJS..264...10K,2023ApJ...946....3K}.   

Of course, $\Upstot$ and $\seff$ can also be empirically measured in observations.  Surveys of hundreds of nearby normal galaxies at $\sim$ kpc scales show that $\Upstot\sim 1000 \kms$ \citep{2008AJ....136.2782L,2017ApJ...835..201H,2020ApJ...892..148S,2021MNRAS.503.3643B,2022ApJ...939..101K,2023ApJ...945L..19S}, which agrees with results obtained from theory and numerical simulations \citepalias{2022ApJ...936..137O}. 
The thermal and turbulent velocity dispersion contributions to $\seff$ are each $\sim 5-10\kms$ \citep[e.g.][]{2009AJ....137.4424T,2011MNRAS.410.1409W,2013ApJ...765..136S,2016AJ....151...15M,2017A&A...607A.106M}; magnetic terms are difficult to measure and less certain, but empirical estimates are overall similar in magnitude to kinetic terms \citep[e.g.][]{Heiles_Troland_2005,Beck_2019}. Thus, observations suggest $\seff \sim 10-30\kms$.  The ratio $\tdep/\tdyn=\Upstot/\seff$ is therefore empirically found to be $\sim 100$ (or at most a factor of 10 lower, under extreme conditions), meaning star formation uses up gas slowly compared to the timescale that is relevant to vertical structure and dynamics of the ISM.  

\citetalias{2022ApJ...936..137O} analyzed a set of TIGRESS simulations \citep{2020ApJ...900...61K} sampling the parameter space of $\Sg$ and stellar+dark matter potential as found in nearby galaxies, in which the emergent $\SSFR$ spans four orders of magnitude ($10^{-4} - 1 \sfrunit$).  While this study was by no means a comprehensive sampling of the complete parameter space of star-forming galaxies -- in particular, only solar metallicity conditions were considered -- these results provide a useful initial calibration of parameters needed for subgrid models of the ISM and star formation in cosmological simulations. In particular, a simple power-law fit of $\Upstot$ to the simulation results produced 
\begin{equation}\label{eq:Upsilon_OK22}
\Upstot = 1028 \kms \left(\frac{\Ptot/k_B}{10^4 \Punit}\right)^{-0.212}
\end{equation}
(see Eq. 25c of \citetalias{2022ApJ...936..137O}).   That is, the feedback yield weakly decreases under conditions of higher pressure -- which correspond to higher mean ISM density.  Physically, this is because under conditions of higher density, (i) radiation is attenuated more in its propagation and therefore its ability to sustain thermal pressure is reduced, and (ii) supernova shocks cool when the swept-up mass is slightly lower, injecting less momentum and therefore producing lower turbulent kinetic and magnetic pressures.  

\citetalias{2022ApJ...936..137O} also found that above $\Ptot/k_B =10^4 \Punit$, the mass weighted effective velocity dispersion follows 
\begin{equation}\label{eq:seff_prfm_va}
\sigma_{\rm eff,avg}
=12 \kms [\Ptot/(10^4 k_B \Punit)]^{0.22}, 
\end{equation}
while at lower $\Ptot$ this mean effective velocity dispersion begins to flatten.
Considering the full set of simulations down to $\Ptot/k_B \sim 10^3 \Punit$ and $n_{\rm H}\sim 0.1\pcc$, \citetalias{2022ApJ...936..137O} also fitted a power-law 
$\Ptot \propto n^{1.43}$ between midplane total pressure and density (see their Eq. 27).   
This fit does not separate out the flattening of $\seff$ at low 
pressure and density.\footnote{Physically, a single 
power law eEoS cannot continue to extremely small values of pressure and density (below the \citetalias{2022ApJ...936..137O} simulated range), because the thermal sound speed of the warm neutral ISM places a floor on $\seff$; using the single power-law fit 
would lead to arbitrarily low values of $\seff$ at low pressure and density. While it is ``safe'' to use a single power law eEoS such as Eq. 27 of \citetalias{2022ApJ...936..137O} at $n_{\rm H}\gtrsim 0.1\pcc$, a (two- or multi-part) fit of $\seff$ that extends to low density and pressure is likely needed in order to cover the full range of galactic conditions, including ultra-diffuse galaxies \citep[e.g.][]{2022ApJ...939..101K}.} 
The midplane pressure-density relation translates to midplane velocity dispersion
\begin{equation}\label{eq:seff_prfm_mp}
    \sigma_{\rm eff,mid}=9.8 \kms [\Ptot/(10^4 k_B \Punit)]^{0.15}.
\end{equation}
With $\Ptot = \seff^2 \rhog$, these are respectively equivalent to eEoS pressure-density relations
\begin{equation}\label{eq:eos_av}
    \log [\Ptot/(10^4 k_B \Punit)]
    = 1.8\log(n_{\rm H}/\pcc) + 0.7\, ,
\end{equation}
for \autoref{eq:seff_prfm_va}
and 
\begin{equation}\label{eq:eos}
    \log[\Ptot/(10^4 k_B \Punit)]
    = 1.43\log(n_{\rm H}/\pcc) + 0.3\, ,
\end{equation}
for \autoref{eq:seff_prfm_mp}.

Together, the calibration of $\Upstot$ and $\seff=\sigma_{\rm eff,avg}$ from TIGRESS simulations in \citetalias{2022ApJ...936..137O} produces $\Upstot/\seff =86 (\Ptot/k_B/10^4)^{-0.43}\propto \rhog^{-0.8}$ for the coefficient in \autoref{eq:tdep_PRFM} (above $\Ptot/k_B =10^4 \Punit$).  When combined with the inverse square-root scaling of $\tdyn$ with density, this implies a significantly steeper scaling of the SFR with density than is typically adopted in cosmological simulations, roughly $\dot m_*/m_\mathrm{g} = 1/\tdep \propto \rhog^{1.3}$ rather than $\propto\rhog^{0.5}$. If instead we use the calibration $\seff=\sigma_{\rm eff,mid}$, we obtain $\Upstot/\seff \propto \rhog^{-0.5} $, i.e. approximately $\dot m_*/m_\mathrm{g}  \propto \rhog$.

As noted above, the calibration in \citetalias{2022ApJ...936..137O} does not include varying metallicity.  At lower metallicity, the feedback yield $\Upstot$ is expected to be higher, both because  radiation propagates more effectively (assuming also reduced dust abundance), and because cooling is reduced, which enhances supernova momentum injection \citep{1998ApJ...500...95T,2019ApJ...881..160B,2020ApJ...896...66K,Steinwandel2020,2023ApJS..264...10K}. If higher density is correlated with lower metallicity as a function of increasing redshift, metallicity-dependent effects would (partly) offset density-dependent effects in $\Upstot$.

The effective velocity dispersion $\seff$ is also likely to depend on metallicity, but this has not yet been characterized numerically, and even the trend of metallicity dependence is difficult to predict theoretically.  We further note that a caveat in using the calibration of $\seff$ from \citetalias{2022ApJ...936..137O} is that the original ``TIGRESS'' simulations analyzed there included supernovae and far-UV heating, but did not include ``early feedback,'' notably the ionizing radiation from short-lived massive stars.  The new, more advanced ``TIGRESS-NCR'' implementation described in \citet{2023ApJS..264...10K} \emph{does} include ionizing radiation, computed via adaptive ray tracing from source star clusters; the radiation pressure force (proportional to the UV flux) is also implemented in TIGRESS-NCR. 
Initial tests show that while  
early feedback does not significantly affect $\Upstot$, it does appear to reduce $\seff$ by $\sim 30\%$ in high-pressure galactic environments \citep{2023ApJ...946....3K,2024arXiv240519227K}. While $\seff$ still increases with $\Ptot$, more comprehensive numerical studies are required for systematic, quantitative assessment and physical understanding of the effects of early 
feedback. 

More generally, if we suppose that $\Upstot=\Upsilon_0(\Ptot/P_0)^{-\alpha}$ and $\seff = \sigma_0 (\Ptot/P_0)^\beta$ for $\Ptot/P_0>1$, the eEoS will be $\Ptot=P_0 (\rhog/\rho_0)^{1/(1-2\beta)}$ for $\rho_0=P_0/\sigma_0^2$, while the depletion time will be $\tdep = (\Upsilon_0/\sigma_0)(\rhog/\rho_0)^{-(\alpha+\beta)/(1-2\beta)}\tdyn$. 
Even in the (unrealistic) case of $\seff$ independent of environment, i.e. $\beta=0$,
$\dot m_*/m_\mathrm{g} = 1/\tdep$ would still increase with density roughly $\propto \rho^{0.5 + \alpha}$ for $\alpha >0$, because $\Upstot$ unambiguously decreases at higher density from radiation attenuation and earlier SNR cooling.  
By running additional TIGRESS-NCR simulations over a range of ISM metallicity and galactic conditions \citep{2024arXiv240519227K}, and fitting the resulting $\Upstot$ and $\seff$, it will be possible to fully calibrate subgrid models for the SFR and the eEoS.   

We note that the PRFM theory predicts a relationship between equilibrium pressure and equilibrium star formation rate. Therefore, the calibrations for the feedback yield, eEoS, and velocity dispersion from TIGRESS simulations are based on fits to the time-averaged values. There exist significant variances in all of these quantities due to the dynamic nature of the star-forming ISM. 
While in principle one might consider sampling from a distribution for $\Upstot$ or $\seff$, the most naive approach to this -- employing independent sampling -- would not improve the representation of the true physical state. This is because the temporal variations of density, pressures (thermal, turbulent, and magnetic), and the star formation rate have complex correlations arising from the interaction of many different physical effects in a high-dimensional system. A time-dependent extension of the PRFM theory would be needed in order to develop subgrid models for star formation and the eEoS that properly represent correlated variations about equilibrium values.  

\subsection{Application to Prediction and Modeling of SFRs}

\begin{figure*}
    \centering
    \includegraphics[scale=0.5]{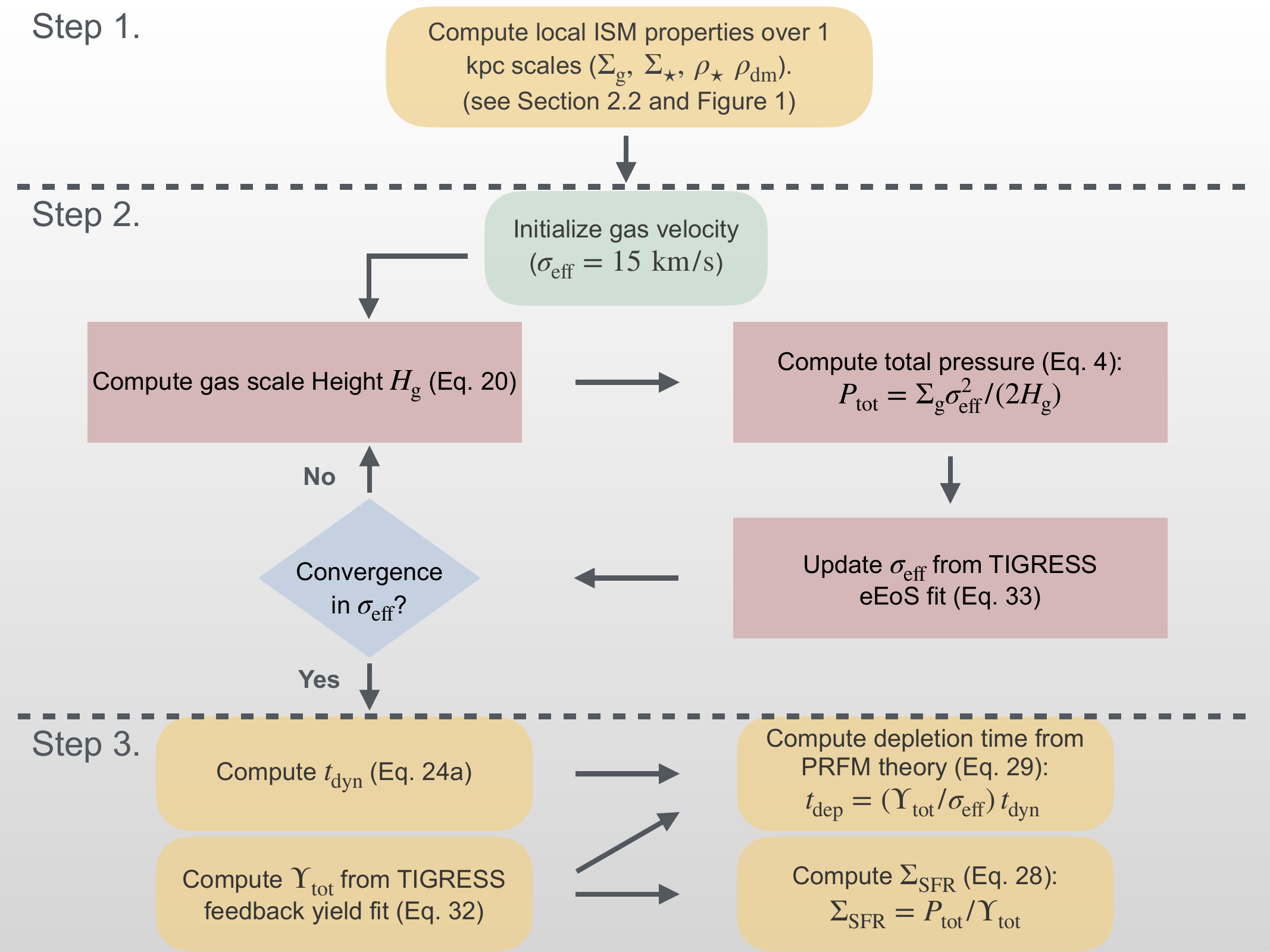}
    \caption{A flowchart displaying a step-by-step summary of the equations used to implement the
PRFM prescription for post-processing the TNG50 simulation outputs. The first step, above the first dotted horizontal line, describes determination of properties from the simulation.  The second step is simultaneous determination of scale height $\Hg$, effective velocity dispersion $\seff$, and total pressure $\Ptot$ assuming vertical equilibrium and a calibrated eEoS.  The  third step, below the second horizontal dotted line, is determination of the SFR from PRFM theory and calibrated feedback yield. For testing how well vertical equilibrium is satisfied within TNG, only the first and third rows of the flowchart are needed, adopting $\seff$ measured from the TNG simulation.}
    \label{fig:flowchart}
\end{figure*}

\REV{The PRFM theory, with calibrations from resolved star-forming ISM simulations, can be used to make predictions for galaxy observations and as a subgrid model for the SFR and eEoS in galaxy formation simulations:
\begin{itemize}

\item[(A)] For \textit{observations where the stellar disk thickness can be directly measured or statistically inferred, and $\seff$ can be estimated from observed linewidths (with an appropriate enhancement for magnetic field)}, an equilibrium estimate of $\Ptot$ is obtained using \autoref{eq:Hg_withdm}.  This $\Ptot$ is then used in a calibrated relationship for $\Upstot$ (\autoref{eq:Upsilon_OK22}), leading to the prediction for star formation rate per unit area $\SSFR=\Ptot/\Upstot$.

\item[(B)] For \textit{galaxy formation simulations where the resolution is high enough for the true thicknesses of the gas and stellar disks to be resolved}, the measured density can be used with a calibrated eEoS (such as \autoref{eq:eos_av}) to set the effective pressure $\Ptot$ of unresolved multiphase ISM gas. Calibrations for $\Upstot$ and $\seff$ as a function of $\Ptot$ (such as those in \autoref{eq:Upsilon_OK22} and \autoref{eq:seff_prfm_va}) can then be used to set the coefficient in \autoref{eq:SFR_PRFM}.  For $\tdyn$,    \autoref{eq:tdyn_rho} may be used.

\item[(C)] For \textit{galaxy formation simulations where the stellar disk thickness is resolved but the gas disk thickness is unresolved}, an equilibrium estimate of $\Ptot$ is obtained using \autoref{eq:Hg_withdm} in \autoref{eq:pressure-weight}. This requires  $\seff$, obtained from an eEoS (here we consider both the eEoS of \citetalias{Springel:2003} and the calibration from TIGRESS given in \autoref{eq:seff_prfm_va}).  The PRFM prediction for SFR in  \autoref{eq:SFR_PRFM} then uses \autoref{eq:tverta} 
for $\tdyn$, and a calibration for $\Upstot$ as a function of $\Ptot$ (here we use \autoref{eq:Upsilon_OK22}). 

\item[(D)] For \textit{galaxy formation simulations where disks are vertically  unresolved}, if it is reasonable to assume that dark matter is unimportant in confining the disk and that $\Hg \approx H_*$, the forms in 
\autoref{eq:simple_tdep} may be adopted, using a calibration such a \autoref{eq:Upsilon_OK22} to evaluate $\Upstot$.  If confinement by dark matter is non-negligible, instead the more general expressions in \autoref{sec:equal_thickness} would be used.

\end{itemize}
 
Of course, we may expect that in the future, inclusion of varying metallicity in ISM simulations will produce generalizations that can be substituted for the calibration relations given here (\autoref{eq:Upsilon_OK22}--\autoref{eq:eos}).  

For the purposes of the present work, in Figure~\ref{fig:flowchart} we provide a flowchart displaying a step-by-step summary of the equations used to implement the PRFM prescription for post-processing the TNG50 outputs. 
Here, we adopt the conservative assumption that the gas disk scale height is unresolved in TNG50, while the stellar disk is resolved \citep{2019MNRAS.490.3196P}, so that case (2) in \autoref{sec:res_usage} is used for $\Hg$, and case (C) above is applied for the PRFM SFR.  
We shall compare the predicted equilibrium pressure and disk thickness with those measured in the simulation to show that the gas disk thickness is marginally resolved.  
}

\begin{figure*}
    \centering
    \includegraphics[width=\linewidth]{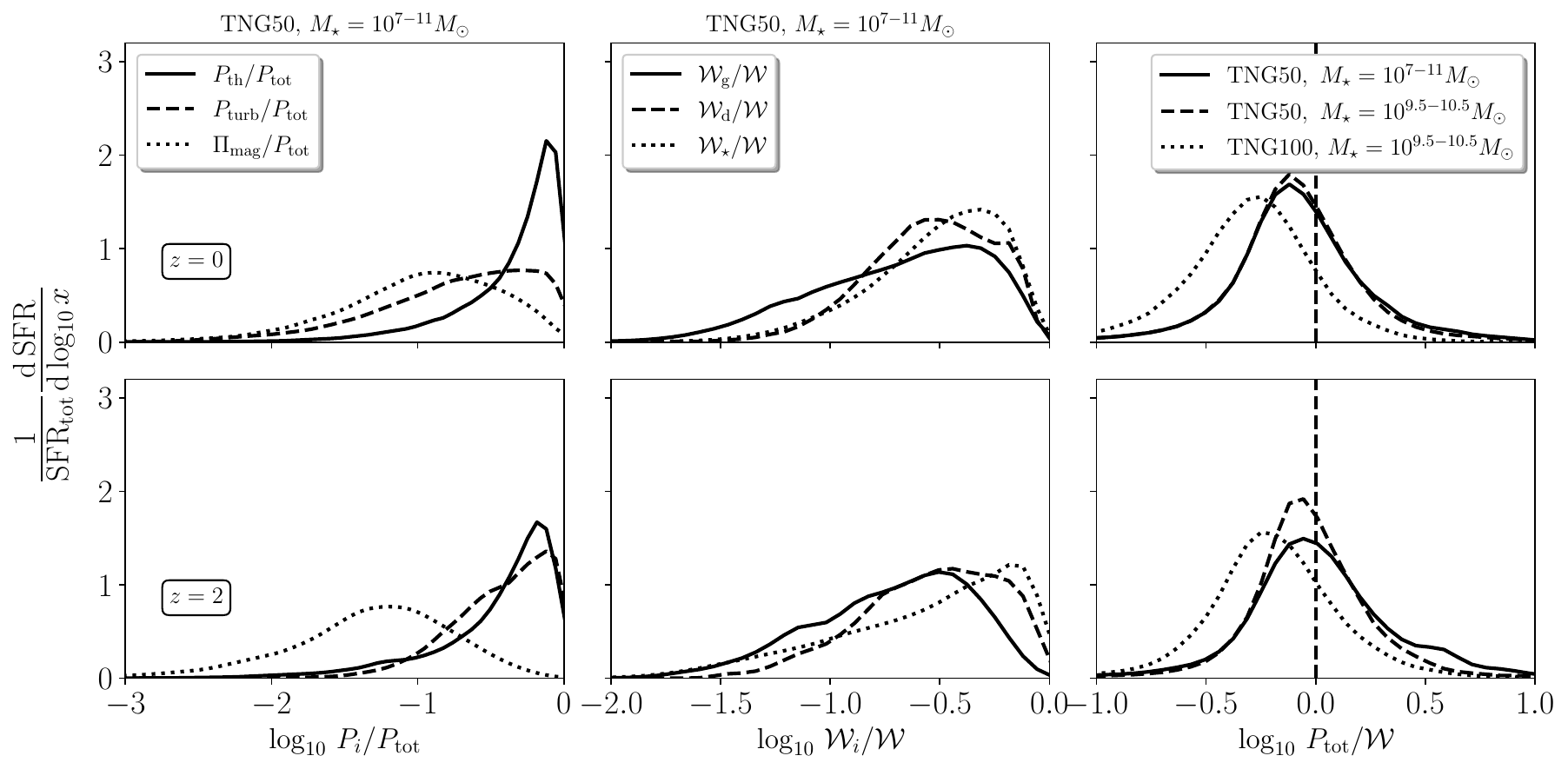}
    \caption{Distributions,  weighted by SFR, of ISM pressure and expected weight as measured in TNG50 
    at $z=0$ (top) and $z=2$ (bottom). For each panel, we show  
    ${\rm SFR}_\mathrm{tot}^{-1} d\,{\rm SFR}/d\log_{10} {\it x}$,     
    where $x$ represents the property on the $x$-axis.  \textit{Left}: Contribution of various pressure components (thermal $P_{\rm th}$, turbulent $P_{\rm turb}$, magnetic $\Pi_{\rm mag}$) to the total pressure $P_{\rm tot}$ as measured at the midplane in TNG50 for the stellar mass range $M_{\star}=10^{7-11}\, M_{\odot}$.
    \textit{Middle}: Contribution of different weight components (gas ${\cal W}_{\rm g}$, star ${\cal W}_{\rm *}$, dark matter ${\cal W}_{\rm d}$) to the total expected weight ${\cal W}$ in TNG50 for $M_{\star}=10^{7-11}$.
    The contribution to expected weight from stellar gravity slightly exceeds the other contributions, particularly at high-$z$ (bottom). 
    \textit{Right}: Distribution of the ratio between the total pressure at the midplane, $P_{\rm tot}$, and the total expected weight, ${\cal W}$ in TNG50 for the whole stellar mass range (solid), and for the  stellar mass range $M_{\star}=10^{9.5-10.5}\, M_{\odot}$ (dashed).  Here we also compare to results from TNG100 for the high end of the mass range (dotted). $P_{\rm th}$ and $P_{\rm turb}$ are the dominant contributors to $P_{\rm tot}$ at $z=0$ and $z=2$, respectively.  
    The distribution of the ratio $P_{\rm tot}$/${\cal W}$ is close to unity (dashed vertical lines) for TNG50 galaxies, indicating that they are generally consistent with vertical equilibrium; the mean and variance are $\log_{10} (P_{\rm tot}/{\cal W})=-0.08\pm 0.34$ and $-0.01\pm 0.36$ for $z=0$ and $z=2$, respectively. Unlike TNG50,  the distribution of the ratio in TGN100 (dotted) is shifted towards lower values ($P_{\rm tot} < {\cal W}$), resulting in $\log_{10} (P_{\rm tot}/{\cal W})=-0.31\pm 0.31$ and $-0.20\pm 0.34$ for $z=0$ and $z=2$, respectively, due to the low resolution.
    }
    \label{fig:p_w_dist}
\end{figure*}

\section{Comparison of ISM and Star Formation Properties}\label{sec:results}

We begin our comparison by testing whether galaxies in TNG50 are consistent with the expectation for vertical equilibrium outlined in \autoref{sec:equil}.  This quantitatively tests whether the simulated galaxy may be considered vertically resolved.   We then compare the velocity dispersion, eEoS, depletion time, and $\SSFR$ as directly measured in our projected maps from TNG50 with the values that would be predicted from the subgrid models discussed in \autoref{sec:prfm}. For convenience, we use the label ``SH03'' to refer to values as measured directly from the simulations, and ``PRFM'' to refer to values obtained through a combination of theory and numerical calibrations from TIGRESS simulations.

\subsection{Testing Vertical Equilibrium}\label{sec:results_equil}

We first verify whether galaxies in TNG50 do, in fact, satisfy the theoretically-predicted equilibrium. This can be examined, as discussed earlier in \autoref{sec:equil},  by comparing the mid-plane pressure (and its components) to the total weight within the 1 proper kpc patches.  It is important to note that we use theoretical values for the weight under vertical equilibrium (see \autoref{sec:equil}), rather than weight calculated using the gravitational force from the simulation, which is subject to gravitational softening.

We measure the different pressure components from the simulations as follows. For the thermal pressure $P_{\rm th}$, we use the density $\rho$ (proper mass density) and internal energy $u$ (thermal energy per unit mass) of gas particles to compute $P_{\rm th}$= $(2/3)\rho u$. The turbulent pressure $P_{\rm turb}$ is computed using $\rho$ and the gas spatial velocity $v_{\rm z}$ in the $z$-direction (perpendicular to the mid-plane) as $P_{\rm turb} = \rho v^{2}_{z}$. Note that while we use the nomenclature ``turbulent pressure,'' this is simply the vertical Reynolds stress term in the momentum equation, where the mean galactic velocity is subtracted from the velocity of any given particle to obtain $v_z$.  The vertical magnetic stress (combining pressure and tension) 
is computed from the magnetic field vector components as 
$\Pi_{\rm mag} = (B^{2}_{x} + B^{2}_{y} - B^{2}_{z})/(8\pi)$. We then compute the mass-weighted average for all these quantities within the midplane (i.e. $\pm$ 100 pc above/below $z=0$ plane) to create their corresponding pressure projected maps. 
We note that the value of $\Pth$ is based on the eEoS adopted in IllustrisTNG (see \autoref{sec:TNG50}) and therefore represents an effective subgrid pressure, rather than being a true thermal pressure obtained via evolution of an internal energy equation with explicit radiative  heating and cooling, and work terms. We also note that given the limited resolution over the scale of the disk (and resulting high numerical dissipation) as well as  the lack of explicit feedback, the turbulent pressure cannot be expected to be as large as it would be in reality.  Nevertheless, our analysis includes a measurement of $P_{\rm turb}$ since all terms in the momentum equation must be combined in order to assess whether the expected vertical equilibrium is satisfied. 

For the different weight components, we directly use the measured local properties from the projected maps described in \S\ref{sec:localprob} in Equations~\ref{eq:gas_W},~\ref{eq:stellar_W}, and ~\ref{eq:dark_W} to obtain ${\cal W}_{\rm g}$, ${\cal W}_{*}$ and ${\cal W}_{\rm d}$, respectively. To compute  ${\cal W}_{\rm g}$ using Equation~\ref{eq:gas_W}, we use the measured $\Sigma_{\rm g}$. For ${\cal W}_{*}$ using Equation~\ref{eq:stellar_W}, we use the measured $\Sigma_{\rm g}$,  $\Sigma_{*}$, and $H_{*}$. For ${\cal W}_{\rm d}$, we assume $\Omega^{2}_{\rm d}= 4 \pi G \rho_{\rm d}$ and 
use the measured $\rho_{\rm d}$ from the local total mass
density around gas particles, as described earlier (see \autoref{sec:localprob}). 
Since our goal is to test TNG50 galaxies against estimates assuming theoretical vertical equilibrium, here we do not use the measured values from the simulated galaxies for $H_{\rm g}$, but rather use the predicted equilibrium $H_{\rm g}$ following Equation~\ref{eq:Hg_withdm},
which includes contributions from gas, stellar, and dark matter gravity. \REV{For testing how well the measured pressure in TNG agrees with the vertical equilibrium prediction,  we use the measured $\sigma_{\rm eff} = \sqrt{P_{\rm tot}/\rho_{\rm g}}$ to compute ${\cal W}_{*}$ and ${\cal W}_{\rm d}$, instead of using the calibration of $\seff$ from TIGRESS}.

We present the distributions of expected weight and measured pressure, and their  comparison, in \autoref{fig:p_w_dist}.   We show $\Sigma_{\rm SFR}$-weighted distributions at redshifts $z=0$ (top) and $z=2$ (bottom).
The contributions from all 1 proper kpc patches of all galaxies combined in TNG50
are shown.  

In the left panels of \autoref{fig:p_w_dist}, we show the relative contribution of various pressure components (thermal $P_{\rm th}$, turbulent $P_{\rm turb}$, magnetic $\Pi_{\rm mag}$) to the total midplane pressure ($P_{\rm tot}$)
in TNG50 for the whole stellar mass range.
We find that the thermal pressure $P_{\rm th}$ has the highest contribution to $P_{\rm tot}$ at $z=0$, whereas the turbulent pressure $P_{\rm turb}$ is higher at high redshift $z=2$. The  magnetic pressure $\Pi_{\rm mag}$ has a minimal contribution to the total pressure $P_{\rm tot}$ throughout. 

In the middle panels of \autoref{fig:p_w_dist}, we show the relative contribution of various weight components (gas ${\cal W}_{\rm g}$, star ${\cal W}_{\rm *}$, dark matter ${\cal W}_{\rm d}$) to the total integrated weight ${\cal W}$ as computed theoretically in TNG50 for the whole stellar mass range.
For $z=0$, it appears that all components (stellar ${\cal W}_{\rm *}$, gas ${\cal W}_{\rm g}$, and dark matter ${\cal W}_{\rm d}$) contribute approximately by the same amount to the total weight ${\cal W}$, though the stellar contribution slightly exceeds other contributions.
By contrast, observed galaxies in the local Universe have lower contribution from ${\cal W}_{\rm d}$: as shown in \autoref{sec:equil}, the ratio ${\cal W}_{\rm *}/{\cal W}_{\rm d} \sim (3/4)\rho_*/\rho_\mathrm{d}$ if $\Hg \sim H_*$, and in the star-forming regions of observed disk galaxies the stellar-to-dark matter density ratio substantially exceeds unity. 
The relatively similar contribution of dark matter to the weight reflects the properties of TNG50 galaxies as shown in \autoref{fig:all_dist}.

In the right panels of \autoref{fig:p_w_dist}, we show the distributions of the ratio between the total measured midplane pressure $P_{\rm tot}$ and the total expected weight ${\cal W}$ in TNG50 for the whole stellar mass range (solid), and for a stellar mass range $M_{\star}=10^{9.5-10.5}M_{\odot}$ (dashed). For this larger mass range ($M_{\star}=10^{9.5-10.5}M_{\odot}$), we show for comparison results from TNG100 (dotted).  In TNG50 for the whole stellar mass range at $z=0$ and $z=2$, the $\Sigma_{\rm SFR}$-weighted mean and standard deviation of $\log_{10} (P_{\rm tot}/{\cal W})$ are $-0.08\pm 0.34$ and $-0.01\pm 0.36$, respectively.  For $z=2$, the peak is close to unity (dashed vertical lines), indicating that equilibrium is satisfied and the vertical scale height is resolved. 
For $z=0$, the peak is slightly below unity, meaning the total pressure is systematically smaller than the expected weight for the majority of regions.  This suggests that at low redshifts the gas scale height is only marginally resolved in TNG50. 
For the higher mass range (which overlaps with that accessible in the TNG100 simulations), the results for TNG50 are similar.  In this higher mass range, we test the impact of resolution by comparing to results from analysis of the TNG100 simulation. 
For TNG100, the distribution of the $P_{\rm tot}/{\cal W}$ ratio is shifted to lower values, resulting in $\log_{10} (P_{\rm tot}/{\cal W})$ of $-0.31\pm 0.31$ and $-0.20\pm 0.34$ at $z=0$ and $z=2$, respectively. This shows that if the resolution is similar (or lower) to that in TNG100, it is not possible to reach a midplane pressure consistent with the theoretically-predicted equilibrium. As a consequence, the midplane density in the simulation would be lower than it should be realistically.

\begin{figure*}
\centering
\includegraphics[width=\linewidth]{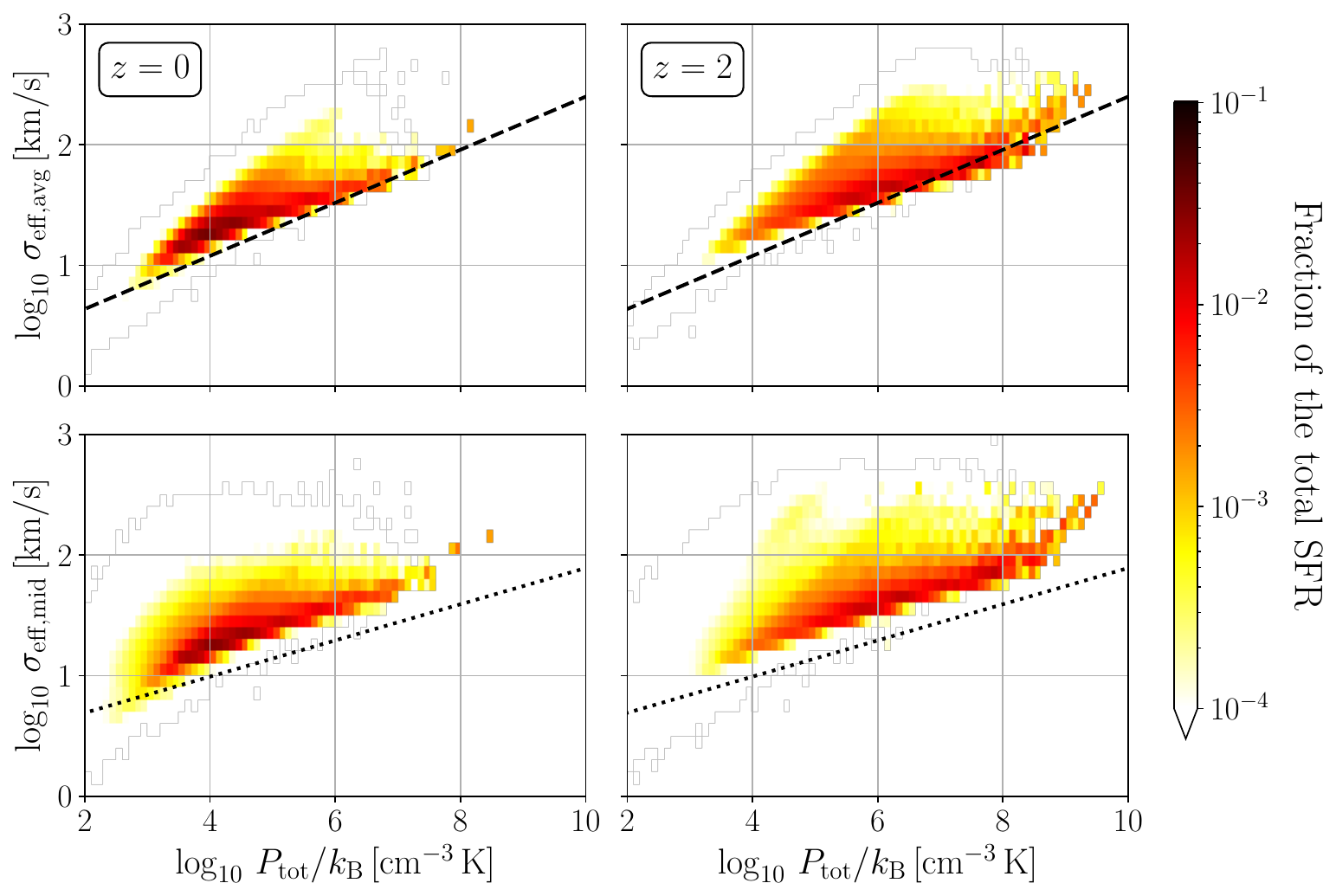}
\caption{The effective velocity dispersion $\sigma_{\rm eff}$ in TNG50 galaxies as a function of the total pressure $P_{\rm tot}$ at $z=0$ (left) and $z=2$ (right). 
The top row shows the mass-weighted vertical average $\sigma_{\rm eff,avg}$, while the bottom row shows the midplane value $\sigma_{\rm eff,mid}$. 
For each panel, the two-dimensional distribution is weighted by the fractional contribution to the total SFR at given redshift from each 1 kpc patch in TNG50 galaxies.  
The TIGRESS fits of the mass-weighted average 
(\autoref{eq:seff_prfm_va}) and the midplane value (\autoref{eq:seff_prfm_mp}) of $\seff$ are shown by the dashed and dotted lines, respectively. The distribution of $\sigma_{\rm eff}$ in TNG50 broadly follows a similar slope and normalization to the results from TIGRESS (especially consistent with $\sigma_{\rm eff, avg}$), albeit with scatter that extends to larger $\seff$.}
    \label{fig:sig_p_dist}
\end{figure*}

\subsection{The Effective Velocity Dispersion of Gas} \label{sec:results_seff}

We now compare the measured effective velocity dispersion of gas from TNG50 with the fits to TIGRESS simulations, as reported in \citetalias{2022ApJ...936..137O}.  For each TNG50 gas particle, $\seff$ is given by $\seff = \sqrt{\Ptot/\rho_{\rm g}}$; the same formula may be used with averaged pressure and density. To compute the mean $\seff$ within each 1 proper kpc patch, we use vertical averages of $\Ptot$ and $\rhog$ within either $z=\pm 10$ kpc or $z=\pm 100$ pc, corresponding to a mass-weighted average $\sigma_{\rm eff,avg}$ or a midplane value $\sigma_{\rm eff,mid}$, respectively. \autoref{fig:sig_p_dist} shows
the two-dimensional histograms of the two measured values of $\seff$ (top and bottom for mass-weighted and midplane values, respectively) as a function of the measured $\Ptot$. 
The histogram is weighted by the contribution from each bin to the total SFR.  Since both pressure and density decline with $|z|$, the ratio  is insensitive to $z_\mathrm{max}$ and the midplane and mass-weighted average values of $\seff$ are similar. 

Also shown in \autoref{fig:sig_p_dist} are the corresponding fitting results as 
presented in \citetalias{2022ApJ...936..137O}, given here in 
\autoref{eq:seff_prfm_va}
for the mass-weighted average, $\sigma_{\rm eff,avg}$, and
\autoref{eq:seff_prfm_mp}
for the midplane value, $\sigma_{\rm eff,mid}$. We note that these fitting functions for $\seff$ represent the time-averaged state over seven TIGRESS models. 
In general, the measured $\seff$ distribution in TNG50 is quite similar in slope and normalization to the mass-weighted average velocity dispersion
from TIGRESS (\autoref{eq:seff_prfm_mp}; dashed). However, there is some scatter in the TNG50 distribution, extending to higher values.  

In the rest of the analysis in this paper, we shall use the TIGRESS fits for the mass-weighted average ${\sigma}_{\rm eff, avg} \propto \Ptot^{0.22}$ fit as a ``theoretical'' value, although we shall drop the subscript ``avg'' for cleaner notation.  
As previously noted, however, $\seff$ remains somewhat uncertain.  It is useful to evaluate how sensitive the predicted gas scale height is to different choices, which we do next.  

\begin{figure*}
    \centering
    \includegraphics[width=\linewidth]{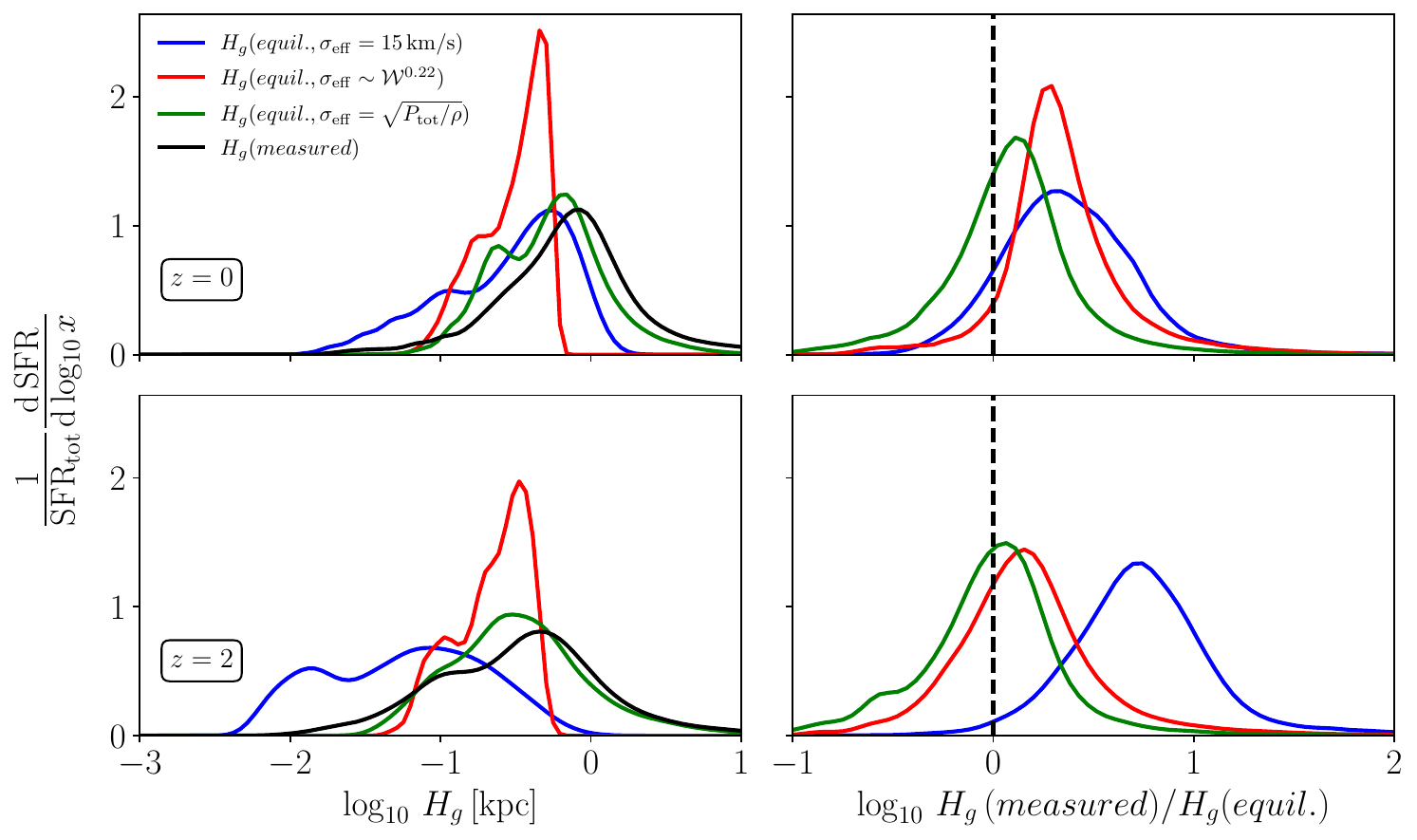}
    \caption{Left: Comparison between the measured scale height $H_{\rm g}(measured)$ in TNG50 (black),  and the predicted equilibrium scale height using different choices for the effective velocity dispersion.  For all predictions, we use \autoref{eq:Hg_withdm}. The different choices of velocity dispersion are:  $\seff = 15 \, {\rm km/s}$ (blue), the TIGRESS fit  
    $\seff \propto {\cal W}^{0.22}$ 
    (i.e., \autoref{eq:seff_prfm_va} with $\Ptot = {\cal W}$; red) for $\cal W$ the ISM weight, and the TNG50 measured value  $\seff=\sqrt{P_{\rm tot}/\rho}$ (green). Right: Ratio of the measured to the predicted equilibrium scale height for each case. The vertical dashed line shows the identity (perfect agreement). The top row shows $z=0$, and the bottom row shows $z=2$. For the theoretical prediction given the measured $\seff$ in TNG50 (green line), $\log_{10} (\Hg(measured)/\Hg(equil.))=0.08\pm 0.34$ and $-0.01\pm0.36$ at $z=0$ and $z=2$, respectively.
    }
    \label{fig:Hg_dist}
\end{figure*}

\subsection{Gas Scale Height}\label{sec:results_Hg}
 
As discussed in \S\ref{sec:equil}, one can use the vertical equilibrium condition ($P_{\rm tot}={\cal W}$) to solve for the predicted gas scale height. Here, for the equilibrium gas scale height $\Hg(equil.)$, we shall use the solution  of the cubic equation (\autoref{eq:Hg_cubic}), which takes into account contributions to the weight from the gravity of the gas, stars, and dark matter. This cubic solution (\autoref{eq:Hg_withdm}) depends on the adopted value for $\seff$, and here we consider different variations of this. These variations include a constant value, a value based on the fit to the TIGRESS simulations (\autoref{eq:seff_prfm_va}), and a value computed directly from TNG50.   The case using a constant, $\seff=15\, {\rm  km/s}$ (motivated by typical measured values in the local Universe), is denoted $H_{g}(equil., \seff=15\, {\rm km/s})$.  The case based on the TIGRESS fit is denoted $H_{g}(equil., \seff
\propto {\cal W}^{0.22})$.
Since in this case, $\seff$ depends on 
$P_{\rm tot} = {\cal W}$ 
(\autoref{eq:seff_prfm_va}), the equilibrium value of $H_{g}$ depends on  $\seff$ (\autoref{eq:Hg_cubic}), and 
${\cal W}= \seff^{2} \Sigma_{\rm g}/2H_{g}$ using the equilibrium value of $H_{g}$, we iteratively solve for $H_{g}$, $\seff$, and ${\cal W}$ assuming an initial value of $\seff=10$ km/s until convergence in $\seff$ is achieved for all 1 kpc patches. We find roughly five iterations are sufficient to achieve convergence.  The case using direct TNG50 measurements is denoted  $H_{g}(equil., \seff=\sqrt{\Ptot/\rho})$, where $\seff$ is computed using the midplane quantities (i.e., the bottom row of \autoref{fig:sig_p_dist}).

In \autoref{fig:Hg_dist}, we present a comparison between these theoretical equilibrium predictions and the measured scale height $\Hg\equiv \Sg/(2 \rhog)$ (for $\rhog$ the midplane value) in terms of the $\Sigma_{\rm SFR}$-weighted distributions at $z=0$ (top) and $z=2$ (bottom).  The left panels show scale heights in proper physical length, and the right panels show ratios between measured and predicted values, with the vertical dashed lines indicating identity (perfect agreement) for reference. As expected, the adopted $\seff$ choice changes the results, potentially dramatically.

First, the figure shows fairly close agreement between the measured and predicted gas height at $z=2$ when using $\seff$ measured from TNG50 \REV{(green line)}. Quantitatively, the $\Sigma_{\rm SFR}$-weighted mean and standard deviation of $\log_{10} (\Hg(measured)/\Hg(equil.,\seff=\sqrt{\Ptot/\rho}))=-0.01\pm 0.36$.
This confirms our earlier conclusion (the bottom right panel of \autoref{fig:p_w_dist}) that TNG50 galaxies at $z=2$ satisfy approximate equilibrium between midplane pressure and the combined weight that is theoretically predicted from gas, stars, and dark matter. Also consistent with our finding that the measured midplane pressure is slightly smaller than the predicted weight at $z=0$ (the top right panel of \autoref{fig:p_w_dist}), here we see that the measured $\Hg$ is systematically larger than the predicted value. The $z=0$ $\Sigma_{\rm SFR}$-weighted mean and standard deviation of $\log_{10} (\Hg(measured)/\Hg(equil.,\seff=\sqrt{\Ptot/\rho}))=0.08\pm 0.34$. It is not surprising that there is a mismatch between the predicted and actual scale height for $z=0$, given that the typical gas cell diameter is $\sim 200 \pc$ for TNG50 \citep{2019MNRAS.490.3196P}, while the predicted median value of the scale height in equilibrium is $\Hg=927$ pc, indicating only marginal numerical resolution.

Second, the bottom row shows that at $z=2$ there is a rough agreement between the peak of the $\Hg$ distribution measured in TNG50 and the prediction based on the $\seff$ fit to the TIGRESS results in \citetalias{2022ApJ...936..137O} (i.e., \autoref{eq:seff_prfm_va}), \REV{as shown with the red line}.  
The agreement is not as close at $z=0$, showing that the TNG50 gas disks are systematically slightly thicker than would be expected if one were to adopt the $\seff$ fit from TIGRESS.
This implies that if a new eEoS calibrated from current high-resolution numerical ISM simulations  (see  \autoref{sec:eos}) were adopted in cosmological simulations, the change in $\Hg$ would be relatively modest at high redshifts. Additionally, it says that the resolution of TNG50 would be sufficient so that the  disk scale heights with a new eEoS would be resolved at $z=2$, although slightly higher resolution would be required for the majority of galaxies at $z=0$.  This suggests that at least at the TNG50 resolution (or slightly higher), it may be relatively straightforward to incorporate more realistic ISM treatments in cosmological simulations, simply by implementing a new eEoS  (see \autoref{sec:eos}), as well as a new star formation rate formulation (see \autoref{sec:tdep}), also calibrated from resolved ISM simulations. 

At resolutions lower than that of TNG50, characteristic of the large-volume cosmological simulations, a different approach  would have to be taken in which the ISM pressure and scale height (and therefore the density) are estimated based on surface densities of stars and gas (which are robust, and independent of resolution provided disks are radially well resolved), as \REV{summarized in cases (2) and (3) in \autoref{sec:res_usage}}. Calibrations of the $\seff$-$\Ptot$ relationship needed for this can be obtained from resolved ISM simulations (e.g., as in \autoref{eq:seff_prfm_va}).

Finally, the results shown in \autoref{fig:Hg_dist} for constant $\seff$ are both interesting and cautionary. At $z=0$, $\Hg$  for $\seff=15$ km/s is not dissimilar to that measured in TNG50, implying that this effective velocity dispersion is a reasonable representation of the actual $\seff$ in TNG50 at low redshift. Indeed, the left panels of \autoref{fig:sig_p_dist} show the peak in the distribution (dark red) near $\seff\approx15\kms$ and $\Ptot/k_B\approx 10^4\Punit$.  However, 
if one adopts $\seff=15$ km/s (blue lines) at high redshift, 
the median predicted scale height at $z=2$ would be an order of magnitude lower than at $z=0$ (left panels). This is a consequence of the higher gravity from denser gas, stars, and dark matter in galaxies at high redshift (\autoref{fig:all_dist}). 
At $z=2$ (the right panels of \autoref{fig:sig_p_dist}), $\seff$ tends to be larger than $15\kms$ and more broadly distributed, without a distinct peak. As a result, the value of $\Hg$ for $\seff=15$ km/s at $z=2$ is much smaller than the TNG50 value. 
Thus, if (motivated by low-redshift observations) one were simply to adopt a constant value of $\seff$ independent of local galactic conditions, it would lead to a scale height much smaller than actually obtained within the TNG50 galaxies at high redshift.  Since the gas density varies inversely with the scale height and the depletion time in TNG50 varies inversely with  the square root of density, this would also have significant consequences for star formation.  

The demonstrated sensitivity of the scale height to the velocity dispersion shows that it is crucial both to obtain proper calibrations for $\seff$ over a wide range of conditions (using resolved ISM simulations or observations) and to implement these calibrations in cosmological simulations.

We note that even when the median value of the measured $\Hg$ agrees with the predicted equilibrium value, there are still variations relative to the equilibrium value ($\sim 0.3-0.5$ dex). Fluctuations about equilibrium are expected in any time-dependent system. Indeed,  as a result of time-varying star formation, feedback, and the thermal and dynamical response to feedback, variations of $\Hg$ of a few tens of percent are evident in the TIGRESS simulations \citep[see, e.g. Fig. 12c of][]{2017ApJ...846..133K}. As mentioned earlier, the $\seff$ adopted for the predicted $\Hg$ represents only temporal averages from seven TIGRESS simulations, which gives rise to the narrow distribution of the predicted $\Hg$ (red). A time-dependent PRFM theory would be needed to fully model the predicted distribution of $\Hg$.
\REV{The distribution of $\Hg$ from PRFM is also cut off more sharply than the $\seff=15\kms$ distribution because no floor has been applied to $\seff$ at low pressure from for this simple comparison. In reality, rather than following \autoref{eq:seff_prfm_va} down to very low values, $\seff$ would have a floor (see footnote 11) at low pressure such that the distribution of $\Hg \propto \seff^2/(\Sg + \Sigma_*)$ would extend to larger values when the gas and stellar surface densities are low.}

\begin{figure*}
    \centering
    \includegraphics[width=\linewidth]{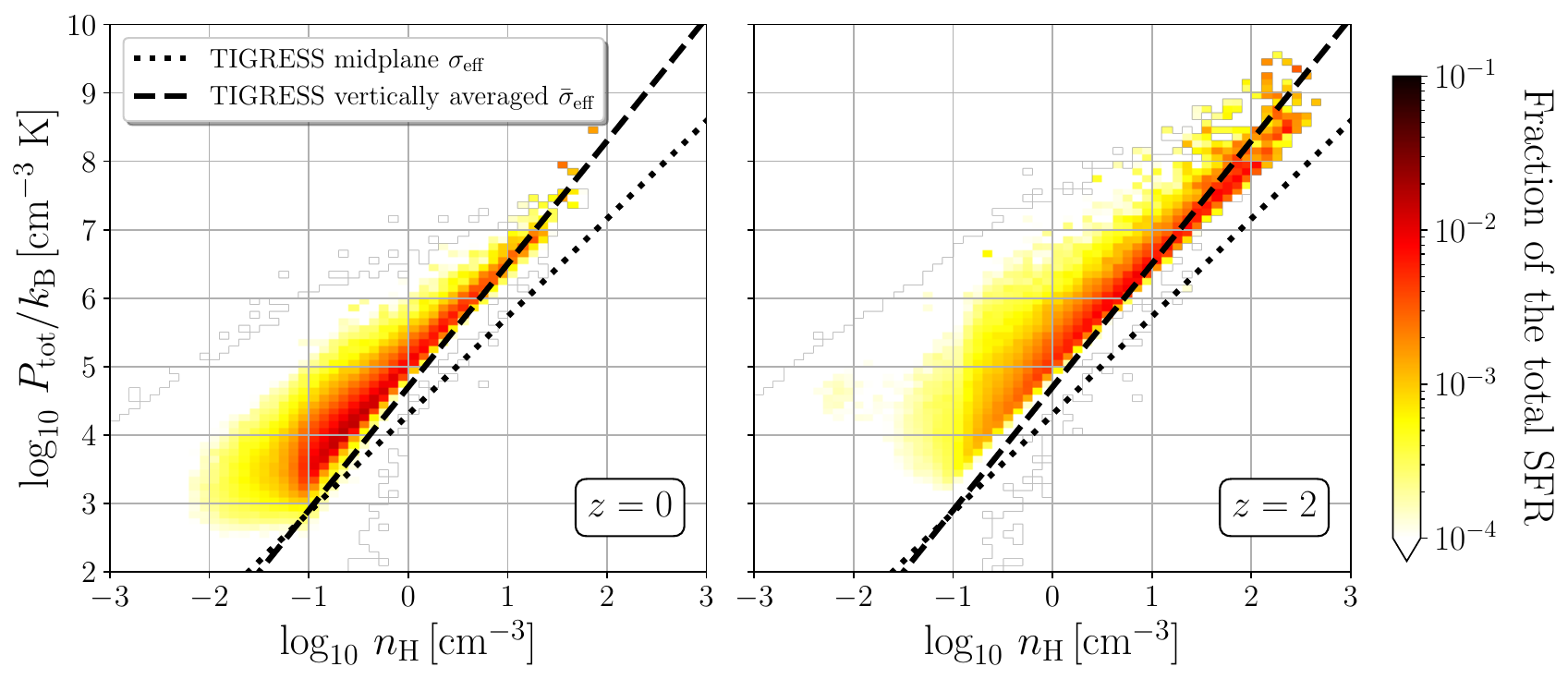}
    \caption{Comparison between the eEoS adopted in TNG50 based on \citetalias{Springel:2003}   (two-dimensional histogram distribution) and the eEoS fit from the TIGRESS simulations analyzed in  \citetalias{2022ApJ...936..137O}  
    (\autoref{eq:eos}, dotted lines, and \autoref{eq:eos_av}, dashed lines).  Both $z=0$ (left) and $z=2$ (right) are shown. The slopes are quite similar, but $P_{\rm tot}$ is shifted upward in TNG50 compared to the TIGRESS fits. }
    \label{fig:eos}
\end{figure*}

\subsection{The Effective Equation of State}\label{sec:eos}

We now turn our attention to the eEoS,
which relates the total pressure to the gas density. Although  characterizing an eEoS is equivalent to characterizing the effective velocity dispersion (since $\Ptot = \rhog \seff^2$)\footnote{
For example, if $P=P_0 (\rho/\rho_0)^{\gamma_{\rm eff}}$ for the eEoS, $\seff=\sigma_0 (P/P_0)^\beta$ for the effective velocity dispersion, with $\beta=(1-1/\gamma_{\rm eff})/2$ and $\sigma_0^2=P_0/\rho_0$.}, 
the eEoS is more directly related to the numerical implementation in a cosmological simulation.  In this context, the eEoS provides an effective pressure that accounts for ``subgrid'' physics that cannot be resolved in the simulation.
In \autoref{fig:eos}, the measured midplane pressure and density in TNG50 are shown as a two-dimensional histogram. The distribution is weighted by the contribution to the total star formation 
at $z=0$ (left) and $z=2$ (right).
For comparison, we plot the TIGRESS fits with dotted and dashed lines, which are equivalent to the midplane value and mass-weighted average of the effective velocity dispersions, respectively. 
These fitting results for $\seff$  expressed as pressure-density relations are
given by \autoref{eq:eos} for the midplane values and 
by \autoref{eq:eos_av} for the mass-weighted average of $\seff$. 

Similar to $\sigma_{\rm eff}$ (\autoref{fig:sig_p_dist}), the 
modified \citetalias{Springel:2003} eEoS adopted in TNG50 (represented by the lower bound in the $\Ptot$ and $n_{\rm H}$ relation) and the fitted eEoS for the midplane values of total pressure and density from TIGRESS (denoted by the dotted lines) are broadly similar in slope. 
The offset to higher pressure in TNG50 compared to the fit from TIGRESS is consistent with the offsets in $\sigma_{\rm eff, mid}$ (the bottom row of \autoref{fig:sig_p_dist}).
The \citetalias{Springel:2003} eEoS is generally more consistent with the pressure-density relation derived by using the mass-weighted average $\seff$ from TIGRESS, which gives a slightly steeper eEoS.

The above TIGRESS calibration does not yet allow for variation in metallicity.  Using an extended TIGRESS-NCR framework \citep{2023ApJS..264...10K,2023ApJ...946....3K,2024arXiv240519227K} where photochemistry is coupled with a UV radiation field obtained by adaptive ray-tracing, the first suite of TIGRESS-NCR simulations with varying metallicities has been developed. 
The results from \citet{2024arXiv240519227K} indicate that the eEoS is very similar for different metallicities (for $Z_{\rm gas}/Z_{\odot}=0.1-1$), but the new TIGRESS-NCR suite gives a slightly shallower power law for the eEoS, with $P_{\rm tot} \propto n_{\rm H}^{1.3}$.

It is worth noting that the implementation of an eEoS \REV{that computes $\Ptot$ as a function of $\rhog$ will only yield a realistic pressure in a cosmological simulation if the resolution is sufficiently high to resolve the disk scale height, since otherwise both $\Ptot$ and $\rhog$ will be lower than they should realistically be (see discussion in  \autoref{sec:res_usage}).  
This is true whether the eEoS is an isothermal relation, or follows an analytic prescription, or is based on calibration from resolved ISM simulations such as TIGRESS. The reason is simply that a realistic density measurement is not possible if the resolution is too low.}  For mass resolution $m_{\rm cell}$, the maximum possible density that can be achieved for a disk with gas surface density $\Sg$ is $\rho_{\rm max} = (\Sg^3/m_{\rm cell})^{1/2}$ (for a monolayer; in practice, if the disk is resolved by a minimum of $N_d$ cells vertically, this would be reduced by a factor $N_d^{-3/2}$).  If $m_{\rm cell}$ is too large, however, this density may be lower than would be expected for the self-regulated state of a star-forming disk of this surface density.  The mean expected density would be equal to the equilibrium pressure divided by $\seff^2$, the square of the effective velocity dispersion.
The equilibrium pressure is discussed in \autoref{sec:equil}; in the simplest case when the gas and stellar disk scale heights are equal and the vertical gravity of the dark matter potential is negligible, from \autoref{eq:H_comb_disk} the equilibrium pressure will be $\pi G \Sg (\Sg + \Sigma_*)/2$.
If $m_{\rm cell}$ exceeds the value given in \autoref{eq:massres}, both the pressure
and the density  would be lower than the equilibrium values should be.   

\REV{If the resolution is too low for physically realistic pressures and densities to be reached within the simulation, an alternative is to simultaneously (1) adopt a formalism for pressure in the simulation that prevents artificial gravitational fragmentation, while (2) obtaining an estimate of what the pressure would be in equilibrium from theoretical considerations (see case (3) in \autoref{sec:res_usage}). The latter estimated pressure is what would be used in setting the SFR via PRFM theory.}

\begin{figure*}
    \centering
    \includegraphics[scale=0.85]{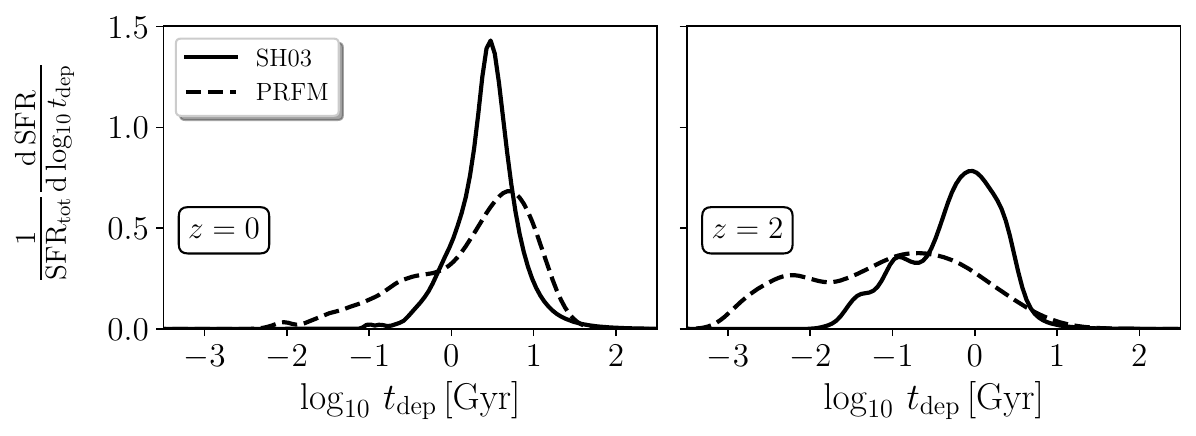}
    \caption{Comparison of the ${\rm SFR}$-weighted distributions of the depletion time ($t_{\rm dep}$) in TNG50 (solid) versus PRFM predictions (dashed) within 1 kpc patches of all galaxies in TNG50 at $z=0$ (left) and $z=2$ (right). While relatively similar at $z=0$, PRFM tends to predict much shorter depletion times at high z, meaning much more efficient star formation.}
    \label{fig:tdep_dist}
\end{figure*}

\begin{figure*}
    \centering
    \includegraphics[width=\linewidth]{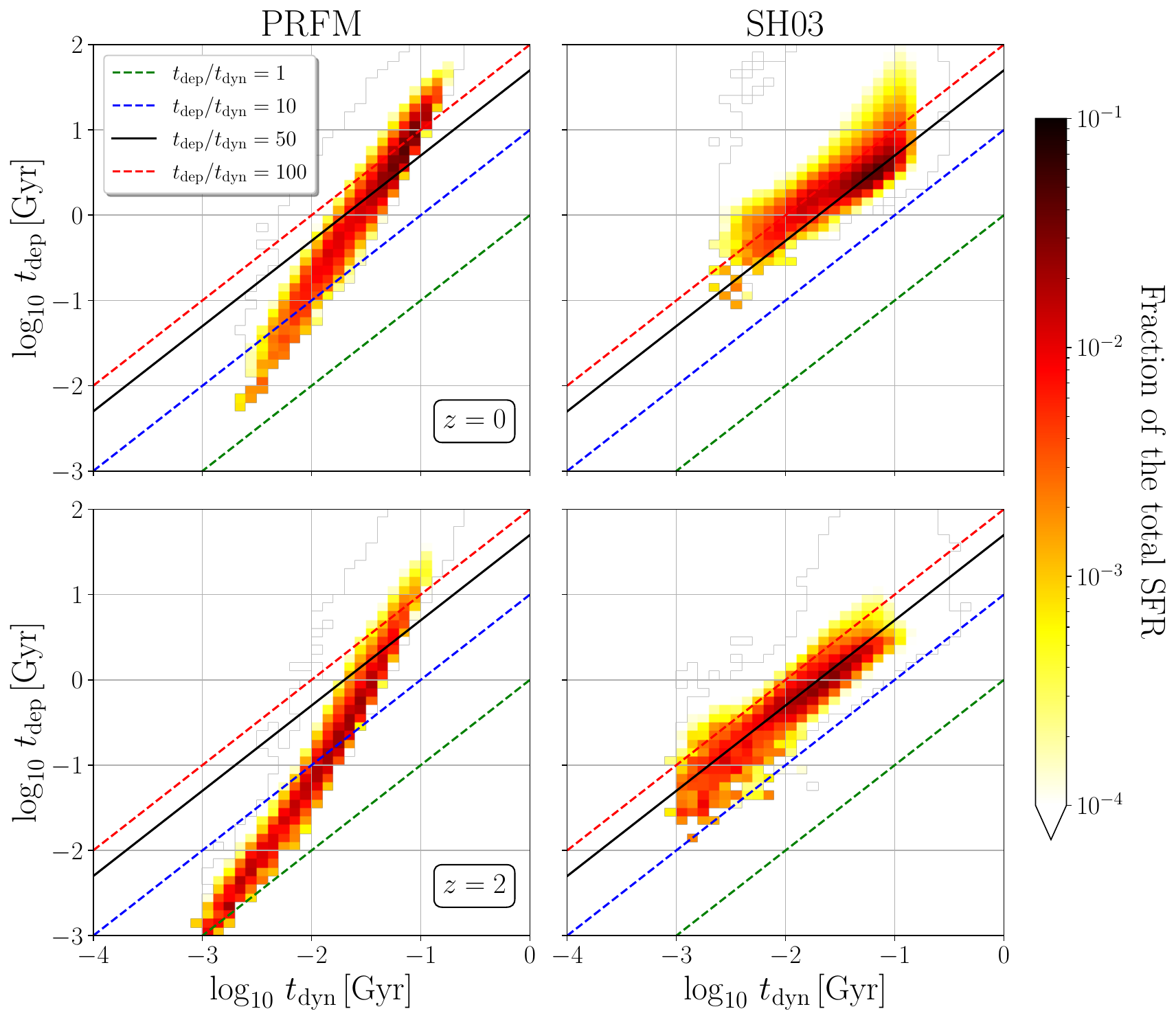}
   \caption{The depletion time $t_{\rm dep}$ versus dynamical time $t_{\rm dyn}$ comparison between the \citetalias{Springel:2003} model adopted in TNG50 (right) and the PRFM model using the calibrations from \citetalias{2022ApJ...936..137O} (left).  Both $z=0$ (top) and $z=2$ (bottom) are shown. The color scale represents the fraction of the total SFR.  For reference, $t_{\rm dep}/t_{\rm dyn}=100$, $10$, and $1$ are shown with dashed lines. The \citetalias{Springel:2003} model follows close to constant $t_{\rm dyn}/t_{\rm dep}$ since the model has been calibrated to a constant star formation efficiency per free-fall time. In the PRFM model, from \autoref{eq:tdep_PRFM}, $\tdep/\tdyn = \Upstot/\seff$; with less efficient feedback (smaller $\Upstot$) and higher velocity dispersion (larger $\seff$) in higher pressure environments where $\tdyn$ is smaller, a reduction in $\tdep/\tdyn$ is expected.  Thus, PRFM predicts a steeper than unity slope for $t_{\rm dep}$ vs. $t_{\rm dyn}$.}
    \label{fig:tdep_tdyn_dist}
\end{figure*}

\begin{figure*}
    \centering
    \includegraphics[width=\linewidth]{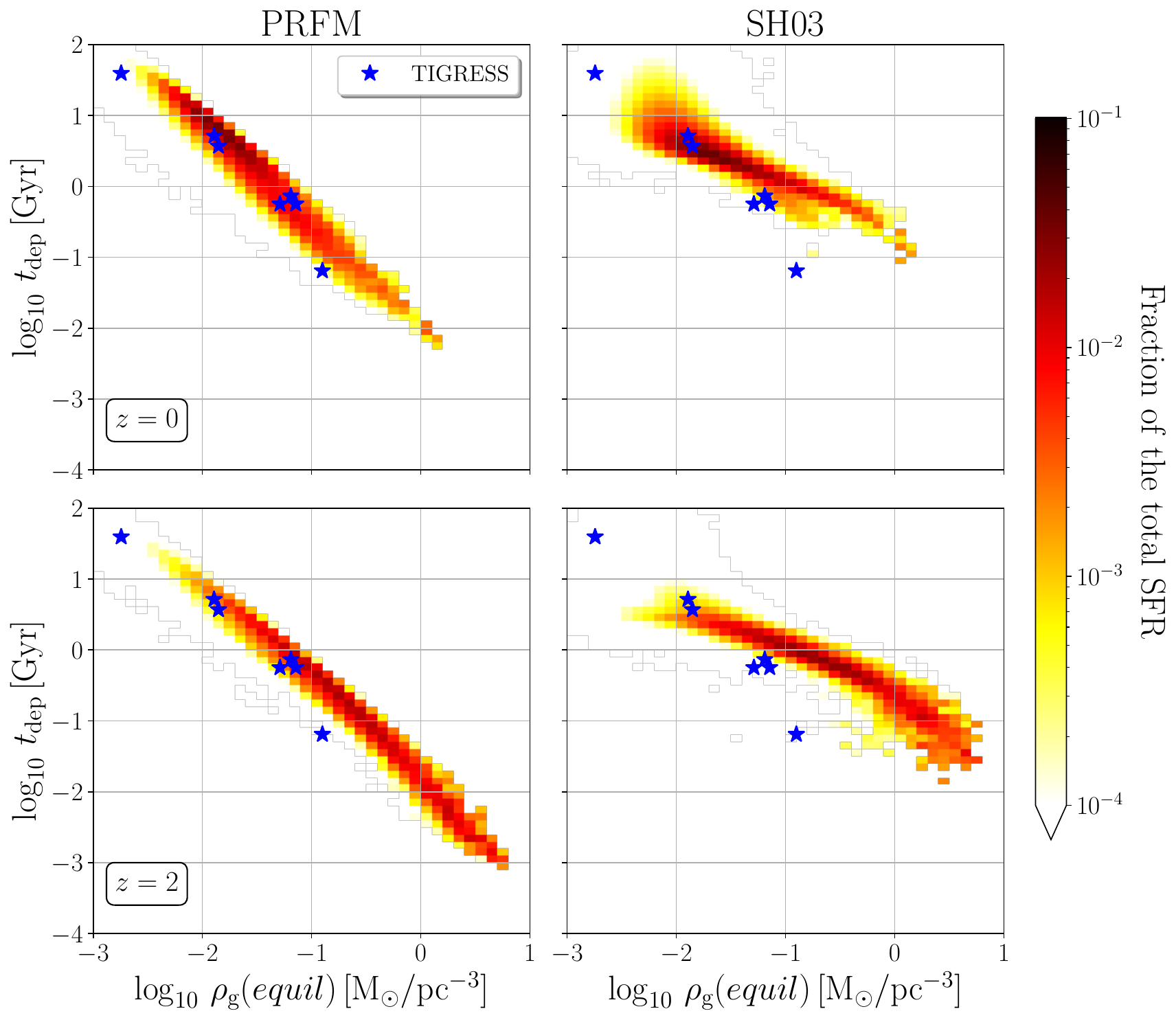}
    \caption{The depletion time ($t_{\rm dep}$) versus the equilibrium gas density $\rhog(equil.) \equiv \Sg/(2\Hg)$, for $\Hg$ in equilibrium obtained using \autoref{eq:Hg_withdm}.  On the right, we show the \citetalias{Springel:2003} value for $\tdep$, while on the left we show the PRFM prediction based on the TIGRESS calibrations in \citetalias{2022ApJ...936..137O}. Both $z=0$ (top) and $z=2$ (bottom) results are shown, with the color scale representing the fraction of the total SFR. Results from TIGRESS runs are also shown with blue stars. Similar to \autoref{fig:tdep_tdyn_dist}, the PRFM model with the TIGRESS calibration predicts a steeper $t_{\rm dep}$ slope than the \citetalias{Springel:2003} model, leading to much shorter depletion times (by nearly an order of magnitude) at high densities.  }
    \label{fig:tdep_rho_dist}
\end{figure*}

\begin{figure*}
    \centering
    \includegraphics[width=\linewidth]{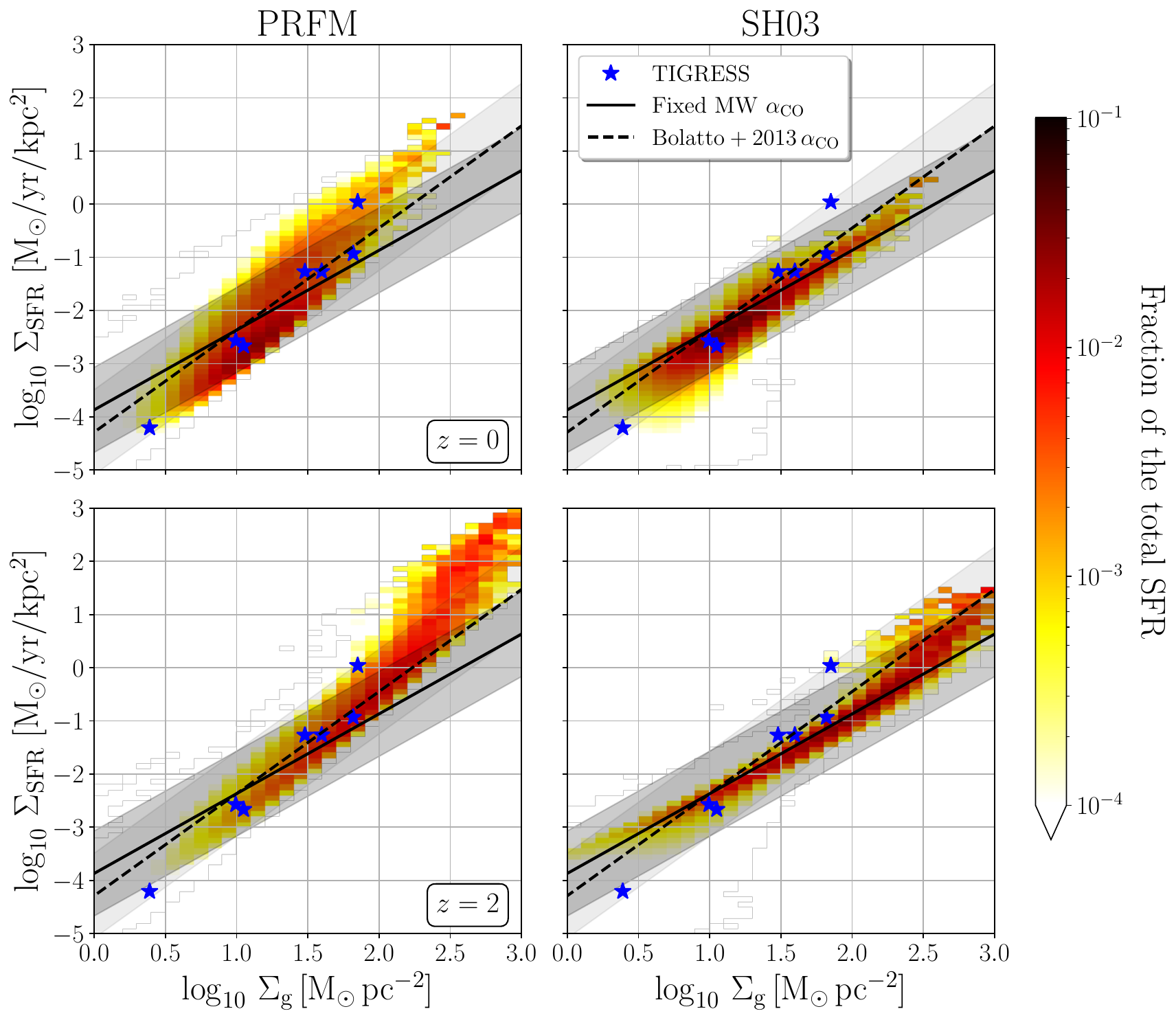}
    \caption{Comparison between Kennicutt-Schmidt relations for the  PRFM model (left) and the \citetalias{Springel:2003} model (right) at $z=0$ (top) and $z=2$ (bottom). The color scale represents contribution to the total SFR. 
    Also shown are results from 
    TIGRESS simulations depicted by blue star symbols, and 
    shaded gray bands 
    showing empirical global KS relations with constant and variable $\alpha_{\rm CO}$,  \citep{2021ApJ...908...61K}; the width of the band represents $\pm 2\sigma$ about the reported best fit power law. 
    The \citetalias{Springel:2003} model follows the constant-$\alpha_{\rm CO}$ empirical global relation, an earlier version of which was used in the original calibration of the model.   
    At high densities, the PRFM model predicts more efficient star formation and hence a steeper slope. }
    \label{fig:KS}
\end{figure*}

\begin{figure*}
    \centering
    \includegraphics[width=\linewidth]{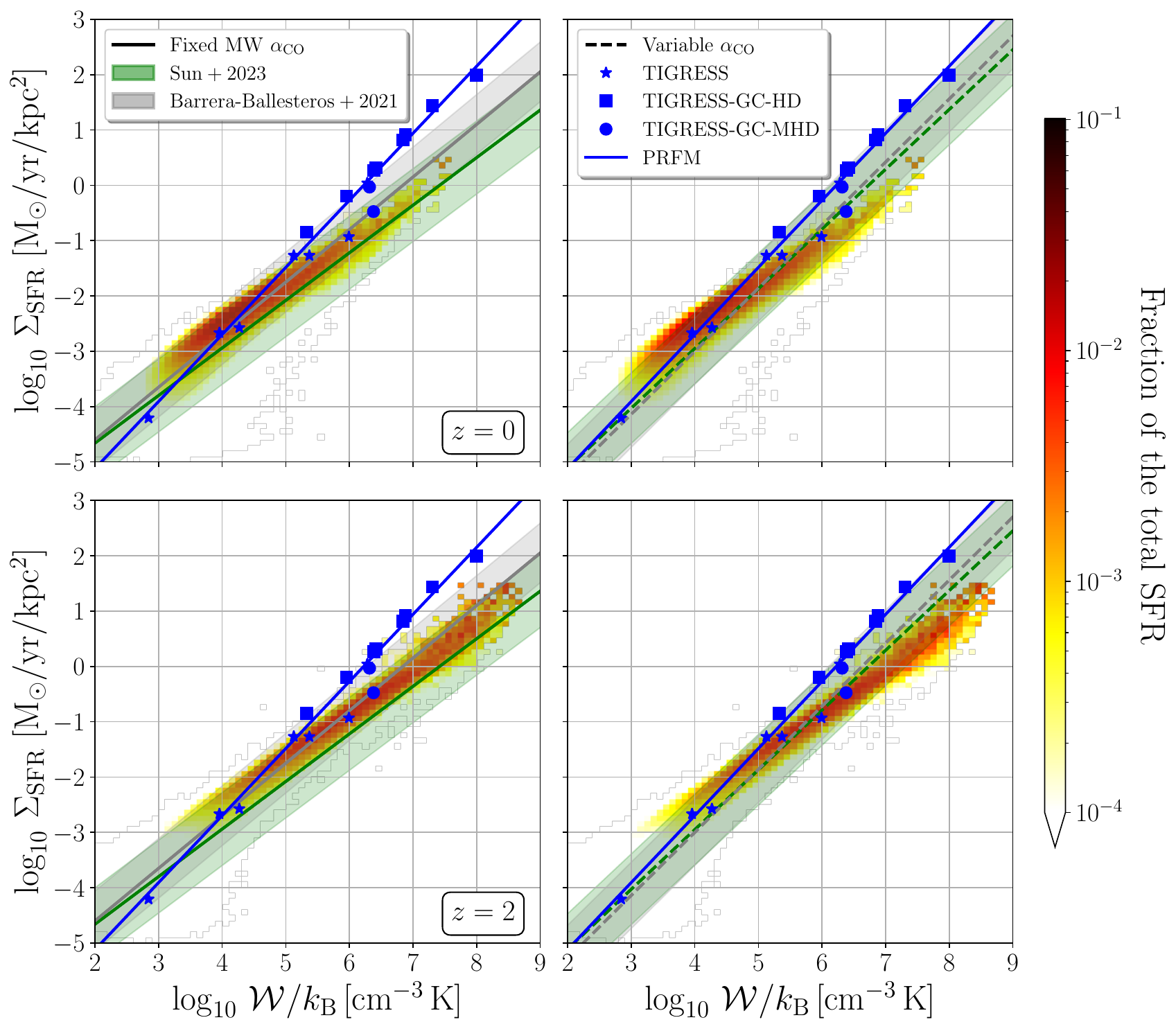}
    \caption{Comparison between models and observations of the relation between the star formation rate surface density  
    $\Sigma_{\rm SFR}$ and the equilibrium total pressure $P_{\rm tot}(equil.)={\cal W}$.  
    The color scale shows the contribution to the total SFR from kpc-scale patches in  TNG50 galaxies based on the native  
    \citetalias{Springel:2003} star formation prescription,  at $z=0$ (top) and $z=2$ (bottom).  Also shown are results from 
    different TIGRESS simulations for local galactic disk models (\citetalias{2022ApJ...936..137O}; blue star symbols) and semi-global galactic center models with (\citealt{2023ApJ...946..114M}; blue circle symbols) and without magnetic fields (\citealt{2021ApJ...914....9M}; blue square symbols). A power-law fit from \citetalias{2022ApJ...936..137O} is shown as blue line. These models are compared to observations of kpc-scale patches in nearby galaxy surveys PHANGS \citep[][green bands]{Sun2023} and EDGE \citep[][gray bands]{BB2021},  adopting both constant (solid, shown to left) and variable (dashed, shown to right) $\alpha_{\rm CO}$. The width of the band represents $\pm 2\sigma$ about the reported best fit power law. 
    }
    \label{fig:SFR_Ptot}
\end{figure*}
\subsection{Gas Depletion Time and Star Formation Relations}\label{sec:tdep}

The depletion time $t_{\rm dep}$, as appearing in \autoref{eq:SFR_SH03}, characterizes the rate of star formation for given gas mass.  TNG50 adopts the 
\citetalias{Springel:2003} prescription for $t_{\rm dep}$ (see \autoref{sec:TNG50}), which depends just on the local gas density in the simulation, with an empirical normalization.  In the PRFM model, $t_{\rm dep}$ in \autoref{eq:tdep_PRFM} depends on gas, stellar, and dark matter density (as appearing in $\tdyn$ in \autoref{eq:tdyn_rho}), and the normalization is set by a theoretical calculation or numerical calibration of feedback yield $\Upstot$ and effective velocity dispersion $\seff$ (see \autoref{sec:calibrations}).  We compare these values of $t_{\rm dep}$ within the 1 kpc pixels from all TNG50 galaxies at $z=0$ and $z=2$. For SH03, we directly use the simulation output to compute $t_{\rm dep}=\Sigma_{\rm g}/\Sigma_{\rm SFR}$.   For PRFM, we use \autoref{eq:tdep_PRFM} \REV{(which instead of the simulation output uses \autoref{eq:yield_def} for $\SSFR$, with the theoretically-computed equilibrium pressure $\Ptot={\cal W}$)}.  For PRFM, the total yield $\Upsilon_{\rm tot}$ has  been calibrated as a function of total pressure from TIGRESS simulations (see \autoref{eq:Upsilon_OK22}); as an input to this we use the predicted equilibrium midplane pressure for the kpc-scale $\Sg$ and the mean stellar and dark matter properties (see \autoref{sec:equil}).  The velocity dispersion $\seff$ is also required for predicting $\Ptot$, $t_{\rm dyn}$, and the star formation efficiency per dynamical time  $\seff/\Upstot$;  for $\seff$ we use the TIGRESS calibration given in  \autoref{eq:seff_prfm_va}. Since the theoretical $\Ptot$ depends on $\seff$, and our $\seff$ calibration is based on the value of $\Ptot$, iteration is required. \REV{As indicated in \autoref{fig:flowchart}}, we iteratively solve for $P_{\rm tot}$ and $\seff$ assuming an initial $\seff=15$ km/s until convergence in all quantities is achieved.  

\autoref{fig:tdep_dist} compares SFR-weighted distributions of $t_{\rm dep}$ from PRFM (dashed) and SH03 (solid) at $z=0$ (left) and $z=2$ (right). At $z=0$, the models have fairly similar $t_{\rm dep}$ distributions (with respective peaks at 3.0 and 5.4 Gyr, and most values between 1 and 10 Gyr).  The PRFM model at $z=0$ does, however, predict $t_{\rm dep}$ extending down to 10$^{-2}$ Gyr, associated with massive star-forming galaxies whose stellar mass $M_{*} > 10^{10} M_{\odot}$.  The range of $t_{\rm dep}$ from the 
\citetalias{Springel:2003} model is smaller, and does not extend below $0.1$ Gyr.   At $z=2$, the differences in the distributions are much larger. The peak of the  SH03 distribution is at $\sim$1 Gyr, whereas the PRFM prediction peaks at $\sim 0.1$ Gyr. Also, the PRFM $t_{\rm dep}$ distribution has a broader range ($\sim 10^{-3}$ to $10^{1}$ Gyr), compared to the range for SH03 (10$^{-2}$ to 10$^{1}$). All else being equal, the shorter $t_{\rm dep}$ predicted by PRFM at $z=2$ implies there would be a higher efficiency of star formation at high redshift if this model were adopted. Of course, since the PRFM model tends to increase the SFR at high pressure and density compared to the TNG50 model, its adoption would in fact alter galactic conditions at high $z$, potentially reducing the gas density within galaxies by driving stronger winds \citep{2020ApJ...900...61K,2020ApJ...903L..34K}.  

We next discuss the reasons behind the difference in the $t_{\rm dep}$ distributions for the two models.  
We first consider the relationship between the dynamical time $\tdyn$ (defined in \autoref{eq:tverta}; see also \autoref{eq:tdyn_rho}) and depletion time $\tdep$. In simple models of star formation, the ratio of $\tdep$ and $\tdyn$  (or more precisely, the free-fall time $\tff$, which typically scales with $\tdyn$) is often assumed to be constant (as in \citetalias{Springel:2003}), but this is not realistic as a general model of star formation on large scales (see discussion and references in \autoref{sec:intro}).  

We present our comparison between the $\tdep$ vs. $\tdyn$ relation based on the PRFM model (left) and based on the modified \citetalias{Springel:2003} model from TNG50 (right) in \autoref{fig:tdep_tdyn_dist} at $z=0$ (top) and $z=2$ (bottom). 
For reference, we also show three different ratios (1, 10, 100) between $\tdep$ and $t_{\rm dyn}$ (dashed lines). The histogram is weighted by the contribution to the total star formation at given $z$.
Evidently, the star formation model based on SH03 follows a nearly linear relation between $\tdep$ and $\tdyn$, corresponding to an efficiency of $\sim 2\%$ (the black solid line, corresponding to $\tdep/\tdyn=50$) from $\tdyn \sim 10^{-3} - 10^{-1}$ Gyr.  This is not surprising, since the \citetalias{Springel:2003} model is based on a fixed efficiency ($\varepsilon_{\rm ff}\sim 0.06$) per $\tff$, and therefore $\tdep/\tdyn=\varepsilon_{\rm ff}^{-1} t_{\rm ff}/\tdyn \sim 20 [1 + (\rho_* + \rho_{\rm d})/\rhog]^{1/2}$. 
The PRFM model, when using the calibrations of $\Upstot$ and $\seff$ from the TIGRESS simulations in \citetalias{2022ApJ...936..137O}, predicts a steeper slope. This is because $\tdep/\tdyn=\Upstot/\seff$, and $\Upstot$ decreases while $\seff$ increases at higher pressure (see \autoref{sec:calibrations}), corresponding to regions with shorter $\tdyn$.  

We caution that the results shown for PRFM at $z=2$ are for illustration only (i.e. they should not be taken as a direct prediction) because the high-$z$ conditions are beyond the regime of the current suite of TIGRESS simulations.  Since the adopted calibration used for the PRFM model has only been tested at higher metallicity and lower pressure, it is not necessarily applicable for $z=2$ galaxies.   Nevertheless, it is quite interesting that  the PRFM model generally predicts shorter $\tdep$ at smaller $\tdyn$ than the \citetalias{Springel:2003} model, even at $z=0$. That is, the PRFM model predicts more efficient star formation at small $\tdyn$.

To explore the models' differences in $\tdep$ as a function of environment, we focus on how $\tdep$ depends on the gas density.
For the gas density, we use the ratio between $\Sigma_{\rm g}$ and the  gas scale height $H_{g}$  from the solution to \autoref{eq:Hg_cubic} with $\seff \propto P^{0.22}$ based on  the calibration from TIGRESS (see \autoref{eq:seff_prfm_va}). We present this comparison in \autoref{fig:tdep_rho_dist}. As before, we use the two-dimensional histogram to show $\tdep$ for the PRFM model (left) and for the \citetalias{Springel:2003} model (right), weighted by the contribution to the total SFR at $z=0$ (top) and $z=2$ (bottom). Data from seven TIGRESS simulation results are plotted by blue star symbols (these are time averages; see Table 2 of \citetalias{2022ApJ...936..137O}). In both cases, $\tdep$ decreases as the gas density increases, since more dense gas tends to form stars more efficiently. As expected based on the results presented in \autoref{fig:tdep_tdyn_dist}, the PRFM model relation is steeper than that of the \citetalias{Springel:2003} model; this is because denser regions correspond to higher pressure regions where the feedback yield is smaller and effective velocity dispersion is larger.  
As expected, the PRFM predictions follow closely data from TIGRESS (blue stars), since the model is calibrated to it. 

We now compare the ``Kennicutt-Schmidt'' (KS)
\citep{1959ApJ...129..243S,1998ApJ...498..541K} 
relations -- meaning the relation between $\Sg$ and 
$\SSFR$ -- for the two models.   
\autoref{fig:KS} presents predictions at $z=0$ (top) and $z=2$ (bottom) for the PRFM (left) and \citetalias{Springel:2003} (right) models. We overlay results from  TIGRESS simulations as blue star symbols.
In \autoref{fig:KS}, we also include shaded bands indicating observed global KS relations reported in~\citet{2021ApJ...908...61K}, for both constant (darker gray) and variable (lighter gray) conversion factor $\alpha_{\rm CO}$ \citep[the respective KS slopes are $1.5$ and $1.9$; see also][for local KS relations with constant or varying $\alpha_{\rm CO}$]{2012MNRAS.421.3127N}.

The \citetalias{Springel:2003} model tracks the observed KS relation for constant $\alpha_{\rm CO}$ at all redshifts since the star formation prescription in \citetalias{Springel:2003} was initially calibrated to reproduce an earlier version of this empirical  relation. 
The PRFM predictions are steeper than the \citetalias{Springel:2003} model, being more consistent with the observed global KS relations for varying $\alpha_{\rm CO}$ where data from TIGRESS exist, and showing a steepening as high surface density.\footnote{Interestingly, very recent observational results employing multi-transition and dust-based calibrations of $\alpha_{\rm CO}$ also show an upturn of the (local) KS relation at high surface densities as found in the centers of nearby galaxies \citep{2023arXiv231016037T}}

The extrapolation shown for the PRFM prediction to the regime of larger $\Sg$ and $\SSFR$ (at $\Sigma_{\rm g}  \geq\,  100 \,M_{\odot}$ pr$^{-2}$) is steeper than that from the  \citetalias{Springel:2003} model.   
As noted above, this suggests the intriguing possibility that implementation of the PRFM model in galaxy formation simulations might produce more efficient star formation especially at high redshift, where densities are higher, than in current cosmological simulations. 
With $\SSFR=\Ptot/\Upstot$ in the PRFM model, and $\Upstot \propto \Ptot^{-\alpha}$ for $\alpha\approx 0.2$ from current TIGRESS calibrations (see \autoref{sec:calibrations}), the implied slope would be $\SSFR \propto [\Sg(\Sg+\Sigma_*)]^{1.2}$ if dark matter is unimportant to the vertical weight and the stellar and gas disks have comparable thickness (see \autoref{eq:simple_tdep}).
However, it should be borne in mind that: (1) the PRFM prediction below/above the range shown for TIGRESS simulations is purely an extrapolation using the same calibrations for $\Upstot$ and $\seff$, (2) the conditions at $z=2$ have lower metallicity that those in the particular TIGRESS models used for the calibration here; in fact, it is expected that $\Upstot$ will be larger at lower $Z$, leading to a reduction in the SFR  \citep[][]{2023ApJS..264...10K,2024arXiv240519227K}.   

Finally, we compare the TNG50 results for the SFR in kpc-scale patches based on the  \citetalias{Springel:2003} prescription to spatially resolved observations in nearby disk galaxies at the same scale.  For TNG50, we plot $\SSFR$ vs. the ISM weight $\cal W$ calculated using \autoref{eq:Hg_withdm}.  The observations come from the PHANGS and EDGE surveys, as presented in \citet{Sun2023} and \citet{BB2021}, respectively, who fit relations between $\SSFR$ and a weight estimate (discussed below  \autoref{eq:approx_weight}). In \autoref{fig:SFR_Ptot}, the fits from both \citet{Sun2023} (green) and \citet{BB2021} (gray) are shown, with both constant (solid) and variable (dashed) $\alpha_{\rm CO}$.  In both top (for $z=0$) and bottom (for $z=2$) panels we overlay results from different TIGRESS simulations, namely \citetalias{2022ApJ...936..137O} (blue star symbols),~\citet{2021ApJ...914....9M} (blue square symbols), and~\citet{2023ApJ...946..114M} (blue circle symbols), and the fit  $\SSFR \propto {\cal W}^{1.2}$  (blue line) from \citetalias{2022ApJ...936..137O}, where the fit is equivalent to the PRFM prediction using \autoref{eq:tdep_PRFM} with \autoref{eq:Upsilon_OK22}. Similar trends to the KS relation comparison (Figure~\ref{fig:KS}) hold.  The  \citetalias{Springel:2003} model is similar, although with a slightly shallower slope, to the empirical results when adopting constant $\alpha_{\rm CO}$.   The PRFM-TIGRESS model is closer to the empirical results when adopting variable $\alpha_{\rm CO}$.

\section{Discussion}\label{sec:conc}

\subsection{Summary of Main Results}

We have presented a detailed comparison between star formation rates and timescales using the native \citetalias{Springel:2003} model, and what would be predicted for the same galactic conditions using the PRFM model with calibrations from the TIGRESS simulations of \citetalias{2022ApJ...936..137O}. In addition, we compare the actual eEoS in TNG with the fit 
presented in \citetalias{2022ApJ...936..137O}.  Results are shown based on averages over 1 proper kpc scales from all galaxies combined in TNG50. \REV{In order to make these comparisons, which call for estimates of the midplane pressure and gas scale height in equilibrium, we derive general formulae for these  quantities in \autoref{sec:equil}.} 

Our key findings are as follows:

\begin{itemize}
    \item For the majority of TNG50 galaxies, the total pressure at the mid-plane $\Ptot$ generally agrees with the total weight ${\cal W}$, with $\Sigma_{\rm SFR}$-weighted mean and standard deviation of $\log_{10} (P_{\rm tot}/{\cal W})$ equal to $-0.08\pm 0.34$ and $-0.01\pm 0.36$ for $z=0$ and $z=2$, respectively. This indicates that TNG50 galaxies are in approximate equilibrium (see right panel of \autoref{fig:p_w_dist}).  However, the measured pressure generally underestimates the weight for lower-resolution TNG100 galaxies.

    \item At the TNG50 resolution, the gas scale height is marginally resolved, in the sense that measured gas scale heights are comparable to vertical equilibrium predictions (see \autoref{fig:Hg_dist}).  The relative resolution is better at high redshift than low redshift.  

    \item While the eEoS in TNG50 has a fairly similar slope to that measured in current TIGRESS simulations of the star-forming ISM \citepalias{2022ApJ...936..137O}, the TNG normalization is a factor of a few higher  (see \autoref{fig:eos}).

    \item The local gas depletion time $\tdep$ relates the SFR to gas mass as 
    $\dot{m}_*\equiv m_\mathrm{g}/\tdep$.  The PRFM model for star formation predicts  a ratio 
    relative to  the vertical dynamical time, $\tdep/\tdyn=\Upstot/\seff$, which decreases for lower $\tdyn$ in high-density regions, because $\Upstot$ decreases and $\seff$ increases at higher pressure.  In contrast, the  \citetalias{Springel:2003} star formation model adopted in the TNG simulation has nearly constant $\tdep/\tdyn$.  As a result, the PRFM model would predict  
    shorter $\tdep$ than in TNG50 in high density regions, where $\tdyn$ is small, particularly at high redshift (see \autoref{fig:tdep_dist}, 
    \autoref{fig:tdep_tdyn_dist}, \autoref{fig:tdep_rho_dist}). 

    \item Both the TNG50 simulation and the PRFM model with TIGRESS calibration predict a SFR for TNG50 galaxies that is consistent with the global empirical Kennicutt-Schmidt  relation ($\Sg$ vs. $\SSFR$) in the range $\Sg \sim 1 - 100\, M_{\odot}\, \rm pc^{-2}$.
    At higher densities, the PRFM model predicts increasingly efficient star formation in comparison to the model adopted in TNG50. The increased efficiency varies inversely with the feedback yield $\Upstot$; since $\Upstot$ decreases in high density environments, this would tend to enhance star formation at high redshift compared to TNG50. 
    The $\Sigma_{\rm SFR}-P_{\rm tot}$ relation based on the \citetalias{Springel:2003} model is similar to, but slightly shallower than, empirical results that adopt constant $\alpha_{\rm CO}$.   The PRFM model is steeper, which is more consistent with the empirical relation when variable  $\alpha_{\rm CO}$ is adopted.
     
\end{itemize}

\subsection{Towards New Subgrid Implementations}
The analysis presented in this paper represents the first step towards implementing new subgrid models for the SFR and the eEoS in 
large-box cosmological galaxy formation simulations; these new models are motivated by the PRFM theory and calibrated from high-resolution TIGRESS simulations of the star-forming ISM. While quantitative results are fairly similar for conditions similar to normal galaxies in the nearby Universe, there are greater disparities under more extreme conditions, especially at high redshift, and this has the potential to significantly alter predictions for galaxy formation over cosmic time. \REV{ In particular, the expected trend would be towards more efficient star formation at high redshift compared to results from current galaxy formation simulations.} We caution, however, that quantitative comparisons at high density and low metallicity will require further calibrations with resolved ISM simulations appropriate for high-$z$ conditions.   In particular, at high redshift the increase of $\Upstot$ at low metallicity would partly offset the decrease in $\Upstot$ at high pressure and density, moderating the enhancement in star formation efficiency (see \autoref{eq:SFR_PRFM}).

\paragraph{PRFM Implementation in Resolved Disks}
For cosmological simulations in which both the gaseous and stellar components of galactic disks are vertically resolved, the implementation of new subgrid models will be relatively straightforward.  This requires: (1) replacing the current adopted eEoS (e.g.~from \citetalias{Springel:2003}) with an eEoS calibrated from resolved star-forming ISM simulations (such as 
\autoref{eq:eos_av}), and (2) replacing current star formation rate prescriptions (e.g.~based on the gas free-fall time and an empirical normalization) with a prescription that employs the generalized vertical dynamical time $\tdyn$ and a normalization based on the feedback yield $\Upstot$ and effective velocity dispersion $\seff$ (see \autoref{eq:SFR_PRFM}).  \REV{Calibrations of the dependence of $\Upstot$ and $\seff$ on pressure from current resolved ISM simulations are presented in \autoref{sec:calibrations}; these will be refined in the future.  Case 1 in \autoref{sec:res_usage} explains how $\tdyn$ would be computed from quantities measured in the simulation in the vertically-resolved  case.} 
We also note that \autoref{eq:massres} gives an estimated mass requirement in a simulation for the gas disk to be resolved vertically.

\paragraph{PRFM Implementation in Unresolved Disks}
A more challenging situation for the implementation of new subgrid models is the case in which the gas disk is not vertically resolved.  In this case, the gas pressure and density values in the simulation are not physically meaningful (they will underestimate the true values), and it is necessary to obtain an estimate of what the pressure should be based on vertical equilibrium considerations.  \REV{Cases 2 and 3 in \autoref{sec:res_usage} explain the steps needed to obtain predicted equilibrium values of $\Hg$, $\tdyn$, and $\Ptot$ from quantities that can be measured robustly. }  
Calibrations from resolved ISM simulations of $\Upstot$ and $\seff$ \REV{in terms of $\Ptot$  would then be used} to obtain the coefficient in the SFR relation of  \autoref{eq:SFR_PRFM}.   

A key challenge \REV{for implementation in cosmological simulations} is to measure on-the-fly the surface densities of gas and stars that enter in the predictions of equilibrium quantities.
For the surface densities, 
a choice must be made regarding the direction for the projection (e.g.~along the direction of the galaxy's angular momentum vector). 
Alternatively, it might be possible to use the ratio between the density and its gradient 
to estimate the surface density.
We leave to future work the exploration of these and other approaches.
Finally, it will be important to test whether models based on quasi-equilibrium assumptions may still be applied in situations such as tidal encounters and mergers where the galaxies strongly disturbed \citepalias[see discussion in][]{2022ApJ...936..137O}.
Careful comparison and testing in idealized simulations of resolved and unresolved disk galaxies, including strongly disturbed systems, will be needed before deploying new methods in cosmological simulations.  While this will take some time, the scientific return will be considerable.

\section*{acknowledgments}
\REV{We are grateful to the referee for helpful comments on improving the manuscript.}
SH acknowledges support for Program number HST-HF2-51507 provided by NASA through a grant from the Space Telescope Science Institute, which is operated by the Association of Universities for Research in Astronomy, incorporated, under NASA contract NAS5-26555. The work of E.C.O. and C.-G.K. was supported by grant 10013948 from  the Simons Foundation to Princeton University, to support the Learning the Universe collaboration. G.L.B. acknowledges support from the NSF (AST-2108470, ACCESS), a NASA TCAN award, and the Simons Foundation. M.C.S. is supported by the Deutsche Forschungsgemeinschaft (DFG, German Research Foundation) under Germany’s Excellence Strategy EXC 2181/1-390900948 (the Heidelberg STRUCTURES Excellence Cluster). We are grateful to the Illustris TNG team for making the TNG50 simulation outputs available to the community for analysis efforts.

\section*{Software}
All codes and data used to produce this work can be found at the following \href{https://github.com/sultan-hassan/tng50-post-processing-prfm}{GitHub Link.}

\bibliography{ref}
\end{document}